\documentclass[twocolumn]{aastex631}

\shorttitle{Molecular Excitation in High-z Quasars}
\shortauthors{Fuxiang Xu}

\newcommand{\cii}{$\rm [CII]_{158 \mu m}\,$}
\newcommand{\ci}{$\rm [CI]_{369 \mu m}\,$}

\newcommand{\fir}{$L_{\rm FIR}\,$}
\newcommand{\tir}{$L_{\rm TIR}\,$}
\newcommand{\nH}{{\rm log}\, (n_{\rm H}/\rm cm^{-3})}
\newcommand{\G}{{\rm log}\, G_0}
\newcommand{\Fx}{{\rm log}\, (F_{\rm X}/\rm erg\, s^{-1}\, cm^{-2})}
\graphicspath{{./}{Figures/}}
\begin{document}

\title{Constraining the excitation of molecular gas in Two Quasar-Starburst Systems at $z \sim 6$ }
\correspondingauthor{Ran Wang}
\email{rwangkiaa@pku.edu.cn,rwangpku@gmail.com}
\author[0000-0003-0754-9795]{Fuxiang Xu}
\affiliation{Department of Astronomy, School of Physics,
Peking University, Beijing 100871, People's Republic of China}
\affiliation{Kavli Institute for Astronomy and Astrophysics,
Peking University, Beijing 100871, People's Republic of China}
\author[0000-0003-4956-5742]{Ran Wang}
\affiliation{Kavli Institute for Astronomy and Astrophysics,
Peking University, Beijing 100871, People's Republic of China}
\author[0000-0002-1815-4839]{Jianan Li}
\affiliation{Department of Astronomy, Tsinghua University,
Beijing 100084, People's Republic of China}
\author[0000-0002-7176-4046]{Roberto Neri}
\affiliation{Institut de Radioastronomie Millimétrique (IRAM), 300 Rue de la Piscine, F-38400 Saint-Martin-d'Héres, France}

\author[0000-0001-9815-4953]{Antonio Pensabene}
\affiliation{Dipartimento di Fisica ``G. Occhialini," Università degli Studi di Milano-Bicocca, Piazza della Scienza 3, I-20126, Milano, Italy}

\author[0000-0002-2662-8803]{Roberto Decarli}
\affiliation{INAF—Osservatorio di Astrofisica e Scienza dello Spazio, via Gobetti 93/3, I-40129 Bologna, Italy}

\author[0000-0002-1478-2598]{Yali Shao}
\affiliation{School of Space and Environment, Beihang University, Beijing, People's Republic of China}

\author[0000-0002-2931-7824]{Eduardo Ba\~nados}
\affiliation{Max Planck Institut f\"ur Astronomie, K\"onigstuhl 17, D-69117, Heidelberg, Germany}

\author[0000-0003-2027-8221]{Pierre Cox}
\affiliation{Sorbonne Université, UPMC Université Paris 6 and CNRS, UMR 7095, Institut d'Astrophysique de Paris, 98bis Boulevard Arago, F-75014 Paris, France}

\author[0000-0002-1707-1775]{Frank Bertoldi}
\affiliation{Argelander-Institut f\"ur Astronomie, University at Bonn, Auf dem H\"ugel 71, 53121 Bonn, Germany}

\author[0000-0002-4227-6035]{Chiara Feruglio}
\affiliation{INAF—Osservatorio Astronomico di Trieste, Via G. Tiepolo 11, I-34143 Trieste, Italy}
\affiliation{IFPU—Institute for Fundamental Physics of the Universe, via Beirut 2, I-34151 Trieste, Italy}

\author[0000-0003-4793-7880]{Fabian Walter}
\affiliation{Max Planck Institut f\"ur Astronomie, K\"onigstuhl 17, D-69117, Heidelberg, Germany}
\affiliation{National Radio Astronomy Observatory, Pete V. Domenici Array Science Center, P.O. Box O, Socorro, NM 87801, USA}

\author[0000-0001-9024-8322]{Bram P. Venemans}
\affiliation{Leiden Observatory, Leiden University, P.O. Box 9513, 2300 RA Leiden, The Netherlands}

\author[0000-0002-4721-3922]{Alain Omont}
\affiliation{Institut d'Astrophysique de Paris, Sorbonne Université, CNRS, UMR 7095, 98 bis bd Arago, F-75014 Paris, France}

\author[0000-0001-9585-1462]{Dominik Riechers}
\affiliation{I. Physikalisches Institut, Universit\"at zu K\"oln, Z\"ulpicher Strasse 77, D-50937 K\"oln, Germany}

\author[0000-0002-1815-4839]{Jeff Wagg}
\affiliation{PIFI Visiting Scientist, Purple Mountain Observatory, No. 8 Yuanhua Road, Qixia District, Nanjing 210034, People’s Republic of China}
\affiliation{World Food Programme, No. 2 Jawatte Ave., Colombo 00500, Sri Lanka}

\author[0000-0001-6459-0669]{Karl M. Menten}
\affiliation{Max-Planck-Institut f\"ur Radioastronomie, Auf dem H\"ugel 69, D-53121 Bonn, Germany}

\author[0000-0003-3310-0131]{Xiaohui Fan}
\affiliation{Steward Observatory, University of Arizona, 933 North Cherry Avenue, Tucson, AZ 85721, USA}

\begin{abstract}
We present NOrthern Extended Millimeter Array observations of CO(8-7), (9-8), and (10-9) lines, as well as the underlying continuum for two far-infrared luminous quasars: SDSS J2054-0005 at $\rm z=6.0389$ and SDSS J0129-0035 at $\rm z=5.7788$. Both quasars were previously detected in CO (2-1) and (6-5) transitions, making them candidates for studying the CO Spectral Line Energy Distribution (SLED) of quasars at $z \sim 6$. Utilizing the radiative transfer code CLOUDY, we fit the CO SLED with two heating mechanisms, including the photo-dissociation region (PDR) and X-ray-dominated region (XDR) for both objects. The CO SLEDs can be fitted by either a dense PDR component with an extremely strong far-ultraviolet radiation field (gas density $ n_{\rm H} \sim 10^6 \, \rm cm^{-3}$ and field strength $G_0 \gtrsim 10^6$) or a two-component model including a PDR and an XDR. However, the line ratios, including \tir and previous \cii and \ci measurements, argue against a very high PDR radiation field strength. Thus, the results prefer a PDR+XDR origin for the CO SLED. The excitation of the high-J CO lines in both objects is likely dominated by the central AGN. We then check the CO (9-8)-to-(6-5) line luminosity ratio $r_{96}$ for all $z \sim 6$ quasars with available CO SLEDs (seven in total) and find that there are no clear correlations between $r_{96}$ and both \fir and the AGN UV luminosities. This further demonstrates the complexity of the CO excitation powered by both the AGN and nuclear star formation in these young quasar host galaxies.
\end{abstract}

\section{Introduction} \label{sec:intro}
Over the past two decades, more than 500 quasars at $z\gtrsim5.3$ have been discovered through optical/Near-infrared (NIR) wide-field multi-band surveys \citep{Fan2000, Jiang2015, Yang2019, Wang2019, Matsuoka2022, Fan2023}. These quasars represent the actively accreting supermassive black holes (SMBHs) at the earliest epoch \citep[ e.g.,][]{Jiang2007, Jiang2016, Shen2019}. A large percent of the high-z quasars were detected in far-infrared (FIR) dust continuum, while most high-z FIR-luminous quasars were detected in various CO emission lines at (sub)millimeter wavelengths, revealing dynamical masses of $\sim 10^{10}-10^{11} \rm \, M_{\odot}$, star formation rates (SFR) of $\sim 100-1000\rm \, M_{\odot}\, yr^{-1}$, and molecular gas reservoirs of $\sim 10^{10}-10^{11} \rm \, M_{\odot}$ within a few kpc scales in their host galaxies \citep[e.g.][]{Bertoldi2003, Riechers2009, Wang2010, Wang2011, Wang2013, Wang2016, Carilli2013, Venemans2017, Venemans2018, Venemans2019, Venemans2020, Pensabene2020, Neeleman2021, Neeleman2023, Walter2022, Shao2022}. 

The interstellar medium (ISM) in galaxies serves as a tool for diagnosing various heating and excitation mechanisms based on its abundant spectral lines and continuum emission from the thermal dust. Recently, (sub)millimeter and radio facilities, such as the Atacama Large Millimeter/sub-millimeter Array (ALMA) and the NOrthern Extended Millimeter Array (NOEMA), have enabled us to measure the multi-phase ISM emission from high-z quasar host galaxies with unprecedented sensitivity and resolution. An essential tracer of the ISM in these galaxies is the \cii emission line, one of the primary coolants of the ISM, originating from the ionized and neutral gas surrounding regions of intense star formation. Notably, observations have revealed significant [CII] emission in $z\sim6$ quasar host galaxies, providing compelling evidence for widespread PDR heating and ongoing star formation activity \citep{Wang2013, Wang2016, Decarli2018, Venemans2012, Venemans2016, Venemans2020, Shao2022}.

The CO emission lines have been used to study the physical conditions in a variety of local and high-z systems \citep[e.g.]{Weiss2005, Weiss2007, Riechers2006, Bradford2009, Bradford2011, Spinoglio2012, Gallerani2014, Rosenberg2015, Wang2019, Yang2019, Li2020a, Li2024}. Low-$J$ (e.g., $J \lesssim 2$) CO emission lines trace the bulk of the molecular gas content. The relations between the line luminosities of different CO transitions and the FIR luminosity are widely discussed which reflects the tight connection between the dense molecular gas and star formation \citep[e.g.,][] {Greve2014, Liu2015, Kamenetzky2016}. The low and mid-$J$ CO transitions (e.g., $3 \lesssim J \lesssim 8$) were observed in samples of high-z submillimeter galaxies and starburst quasar hosts which trace the PDR region powered by the ultraviolet (UV) photons from newly formed high-mass stars \citep[e.g.,][]{Bothwell2013, Yang2017, Decarli2022}. Bright high-$J$ (e.g., $J \gtrsim 9$) CO transitions were detected in quasar host galaxies from local to the highest redshift, suggesting highly-excited molecular gas associated with the powerful AGN activity \citep{Bradford2003, Spinoglio2012, Meijerink2013, Gallerani2014, Li2020a, Pensabene2021}. 

By integrating observational data with modeling techniques, we can effectively determine representative values for crucial ISM physical parameters, including gas kinetic temperature ($T_{\rm kin}$), gas density ($n_{\rm H}$), and interstellar radiation field intensity (UV intensity $G_0$ from PDRs or X-rays intensity $F_X$ from AGN). A common approach employs the Large Velocity Gradient (LVG) model for single/multiple-component fitting of CO Spectral Line Energy Distributions (SLEDs) \citep{Weiss2005, Weiss2007, Riechers2009, Spinoglio2012, Wang2019, Yang2019, Li2020a}. This method has unveiled the properties of molecular gas through a combination of cold ($T_{\rm kin} \sim 20-100 \, \rm K$), low-density ($n_{\rm H} \sim 10^3 - 10^4 \rm \, cm^{-3}$), and hot ($T_{\rm kin} \sim 100-300 \, \rm K$), high-density  ($n_{\rm H} \sim 10^4 - 10^{5} \rm \, cm^{-3}$) components. Radiation transfer models for PDRs and XDRs are explored to further constrain the excitation power and mechanisms. For example, Studies of the low and mid-J CO transitions in the starburst systems reveal PDR conditions with a typical interstellar radiation field strength of $G_0 \sim 10^2-10^3$ and $n_{\rm H} \sim 10^2-10^4 \rm \, cm^{-3}$ \citep{VanDerWerf2010, Meijerink2011, Spinoglio2012, Greve2014, Rosenberg2015, Liu2015}. In comparison, high-J CO emission in high-z quasars requires an extra, highly dense ($n_{\rm H} \sim 10^4-10^5 \rm \, cm^{-3}$) XDR component or other possible scenarios such as mechanical heating, cosmic rays, and shocks \citep{Meijerink2007, Spinoglio2012, Gallerani2014, Pensabene2021, Decarli2022, Li2024}. This highlights the potential contribution of AGN to ISM heating, particularly in the early Universe. 

The studies of CO SLEDs for $z \sim 6$ quasars have been mainly focused on the most optically luminous ($M_{1450} \lesssim -26$) objects so far. To better address the impact of AGN on the nuclear ISM, it is essential to expand the study to less UV luminous and FIR bright systems. In this work, we analyzed the CO SLED in the host galaxies of two FIR luminous quasars, SDSS J2054-0005 at z=6.0379 and SDSS J0129-0035 at z=5.7787. 
These two objects are fainter in the optical \citep[$M_{1450}$=-26.1 and -24.4 mag,][]{Jiang2016} compared to quasars at $z\sim 6$ from previous CO SLED studies \citep{Gallerani2014, Wang2019, Yang2019, Li2020a, Pensabene2021, Li2024}. Especially J0129-0035 is about two magnitudes fainter than the previously faintest one, J1429+5447 \citep{Li2024}. The two quasars were detected in the dust continuum, as well as in the \cii, CO(6-5), and CO(2-1) line \citep{Wang2013, Wang2019a, Shao2019}. The derived FIR luminosities (integrated over the range $42.5$ to $122.5\, \mu\mathrm{m}$) are about $8\times 10^{12}\, \rm L_{\odot}$ for J2054-0005 and $5\times 10^{12}\, \rm L_{\odot}$ for J0129-0035, placing them among the most FIR luminous objects at $z \sim 6$. The ALMA continuum and \cii images reveal compact nuclear star formation on scales of a few kpc \citep{Wang2013, Wang2019a}. Thus, they provide ideal targets for studying ISM excitation in environments with intense star formation and less luminous AGNs. In this paper, we present new observations of CO(8-7), CO(9-8), and CO(10-9) lines, as well as the dust continuum obtained with NOEMA toward J2054-0005 and J0129-0035. The goal is to investigate and further constrain the molecular gas excitation mechanisms in their nuclear regions and explore the dust heating and star formation activity. 

The paper is organized as follows: In Section \ref{sec: obs}, we describe the sample, the observations, and the data reduction. In Section \ref{sec: res}, we present our results. In Section \ref{sec: dis}, we first analyze the dust temperature, mass, and SFR for J2054-0005 and J0129-0035 through SED fitting of the dust continuum. We then use the $\mathsf{CLOUDY}$ radiative-transfer code \citep[version c.17.01, ][]{Ferland2017} to investigate the physical conditions of the molecular gas through an analysis of the CO SLED. Finally, we compare the CO excitation of J2054-0005 and J0129-0035 with local and high-z AGN and galaxy samples. In Section \ref{sec: sum}, we summarize our results.

Throughout this work, we adopt a $\Lambda$CDM cosmology with $H_0 = 70 \rm \; km\, s^{-1}\, Mpc^{-1}$, $\Omega_{m}$=0.3, and $\Omega_{\Lambda}$=0.7.

\section{Observations} \label{sec: obs}
We carried out observations of the CO(8-7) ($\nu_{\rm rest}$= 921.80 GHz), CO(9-8) ($\nu_{\rm rest}$ = 1036.9 GHz), CO(10–9) ($\nu_{\rm rest}$ = 1152.0 GHz) emission lines, together with the underlying continuum from SDSS J2054-0005 and SDSS J0129-0035 using NOEMA (Project ID: S21DD). The observations were carried out from September to November 2021, using 9 to 10 antennas in C/D configurations. We used the Band 2 receiver and the PolyFix correlator covering the CO lines in two setups. In one setup, we observe the CO (8-7) line in the lower sideband (LSB) and the CO (9-8) line in the upper sideband (USB), and in the other setup, we put the CO (10-9) line in the USB and CO (9-8) in the LSB. Each setup covers a total bandwidth of 7.74 GHz and a channel width of 2 MHz. MWC349 was used as the flux calibrator for both J2054-0005 and J0129-0035. The phase/amplitude calibrators were J2059+034 for J2054-0005 and 0106+013 for J0129-0035. The typical calibration uncertainty is $< 10 \%$ in Band 2.

Data reduction was performed using the Grenoble Image and Line Data Analysis System software \citep[GILDAS; ][]{Guilloteau2000} packages CLIC and MAPPING. We extracted the continuum from all channels free of spectral lines in the uv plane with UV$\_$AVERAGE. The uv table for spectral lines was produced via UV$\_$SUBTRACT with the subtraction of the underlying continuum. The uv table of the continuum and spectral lines underwent cleaning with the HOGBOM algorithm, while NATURAL weighting was applied to achieve the maximum signal-to-noise ratio (S/N). The CO(9-8) line data from two tunings were combined to improve the S/N ratio. The processed data were binned to spectral resolutions of 64 km/s to 77 km/s to optimize the sensitivity. The details are listed in Table \ref{tab: obs}.

\begin{deluxetable}{@{\hspace{0pt}}ChCCCChh@{\hspace{0pt}}}
\tablecaption{Summary of the observations \label{tab: obs}}
\tablewidth{0pt}
\tablehead{
\colhead{Line} & \nocolhead{Project ID} & \colhead{$t_{\rm on\_source}$} & \colhead{$n_{\rm ant}$} & \colhead{Beam} & \colhead{rms (channel width)}  & \nocolhead{Facilities} & \nocolhead{References} \\
\colhead{} & \nocolhead{} & \colhead{hrs} & \colhead{} &  \colhead{$\rm arcsec^2$} &  \colhead{mJy/Beam ($\rm km \, s^{-1}$)}  & \nocolhead{} & \nocolhead{}
}
\decimalcolnumbers
\startdata
\rm \textbf{J2054-0005} &&&&&&& \\
\rm CO(8-7) &  S21DD & 7.89 & 9-10  & 3.78 \times 3.01 & 0.31(64) & \rm NOEMA &  \\
\rm CO(9-8) &  S21DD & 15.89 & 9-10  & 2.41 \times 1.06 & 0.26(65) & \rm NOEMA &  \\
\rm CO(10-9) &  S21DD & 8.00 & 9-10  & 2.91 \times 1.57 & 0.36(77) & \rm NOEMA &  \\
\hline
\rm \textbf{J0129-0035} &&&&&&& \\
\rm CO(8-7) &  S21DD & 6.81 & 9-10  & 4.73 \times 2.46 & 0.31(66) & \rm NOEMA &  \\
\rm CO(9-8) &  S21DD & 11.41 & 9-10  & 3.43 \times 2.54 & 0.27(63) & \rm NOEMA &  \\
\rm CO(10-9) &  S21DD & 4.60 & 9-10  & 3.23 \times 2.52 & 0.59(74) & \rm NOEMA &  \\
\enddata
\tablecomments{Columns 1: Line's name. Columns 2: On-source integration time. Column 3: Number of antennas. Column 4: Beam size in FWHM; Column 5: the rms and the corresponding Channel width.}
\end{deluxetable}

\begin{figure*}[htbp]
\plotone{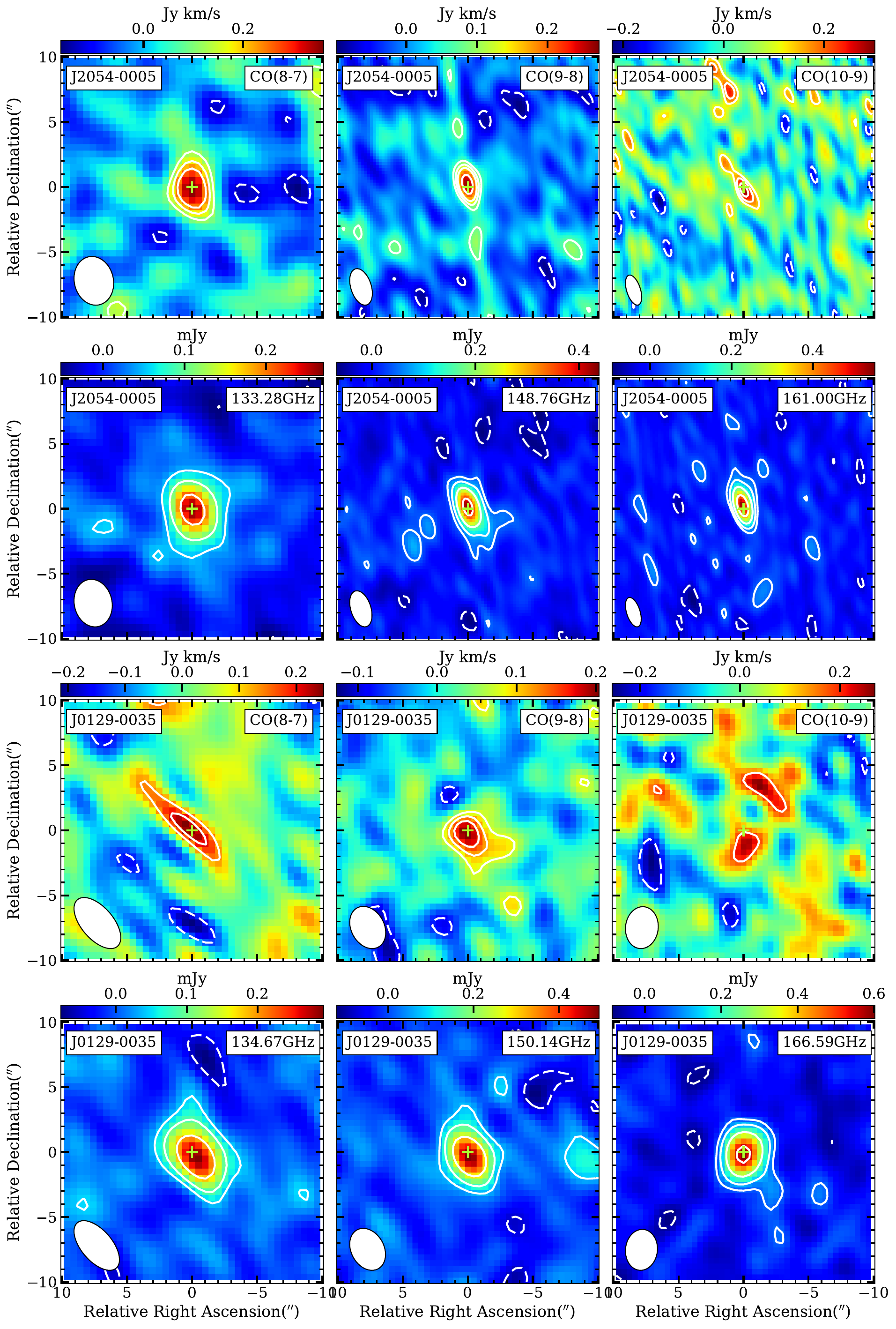}
\caption{The line intensity maps of CO(8–7), (9–8), (10–9) (from left to right) and the corresponding continuum of two objects observed by NOEMA. The top two row panels are for J2054-0005, and the bottom two rows are for J0129-0035. The green-yellow cross on the line intensity map and continuum represents the optical position. The shaded white ellipse in the bottom-left corner of each panel shows the FWHM of the beam. The contours denote the line map contours to [-2, 2, 3, 4, 6] $\times \sigma$ and the continuum contours to [-2, 2, 4, 8, 16] $\times \sigma$}.\label{fig: noema_image}
\end{figure*}

\begin{figure*}[htbp]
\plotone{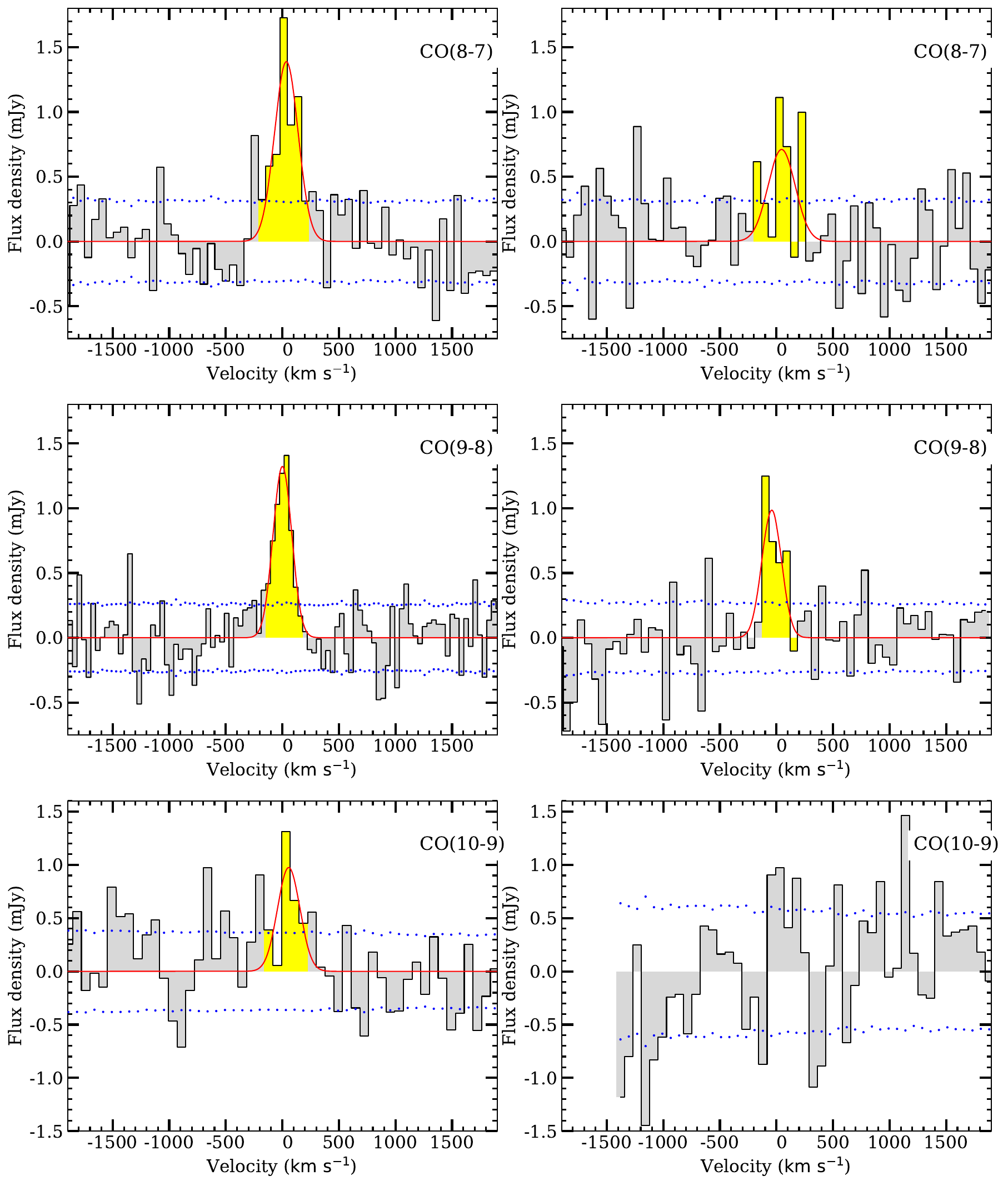}
\caption{The spectra of CO(8–7), (9–8), (10–9) (from top to bottom) of two objects (left: J2054-0005; right: J0129-0035) and the fitting results by one Gaussian mode. The dotted histograms mark the $ \pm 1\sigma$ range in the rms noise. The red lines show the best-fit model of the spectra. The yellow bins represent the velocity range within $ 1.5 \times \text{FWHM}$ of the detected line.} \label{fig: spectral}
\end{figure*}

\begin{deluxetable*}{CCCCChChCC}
\tablecaption{Spectral Line Observations \label{tab: line}}
\tablewidth{0pt}
\tablehead{
\colhead{Line} & \colhead{$z_{line}$} & \colhead{FWHM} & \colhead{Flux} & \colhead{Beam Size} & \nocolhead{Position anlge} & \colhead{Luminosity} &  \nocolhead{rms} & \colhead{Facilities} & \colhead{References} \\
\colhead{} & \colhead{} & \colhead{km s$^{-1}$} & \colhead{Jy km$^{-1}$} &  \colhead{(arcsec)} &  \nocolhead{(degree)} & \colhead{$10^{8}\rm L_{\odot}$} & \nocolhead{mJy/Beam} & \colhead{} & \colhead{}
}
\decimalcolnumbers
\startdata
\rm \textbf{J2054-0005} &&&&&&&&& \\
\rm [C II] &  6.0389 \pm 0.0001 & 236 \pm 12 & 3.23 \pm 0.14  & 0.15 \times 0.11 &  & 30.8 \pm 1.4 &  & \rm ALMA & V20 \\
\rm [C I] &  6.0397 \pm 0.0003 & 487 \pm 115 & < 0.261  & 4.0 \times 2.6  &  & <1.1 &  & \rm NOEMA & D22 \\
\rm CO(2-1) &  6.0394 \pm 0.0004 & 270 \pm 47 & 0.06 \pm 0.01  & 2.42 \times 2.08 &  & 0.069 \pm 0.012 &  & \rm VLA & S19 \\
\rm CO(6-5) &  6.0379 \pm 0.0022 & 360 \pm 110 & 0.34 \pm 0.07  & 5.91 \times 5.12 &  & 1.17 \pm 0.24 &  & \rm PdBI & W10 \\
\rm CO(6-5) &  6.0391 \pm 0.0002 & 229 \pm 20 & 0.29 \pm 0.05  & 0.39 \times 0.30 & -72.736 & 1.00 \pm 0.20 &  & \rm ALMA &  T24a\\
\rm CO(7-6) &  6.0397 \pm 0.0003 & 487 \pm 115 & 0.235 \pm 0.08  & 4.0 \times 2.6  &  & 0.95 \pm 0.38 &  & \rm NOEMA & D22 \\
\rm CO(8-7) &  6.0396 \pm 0.0005 & 246 \pm 29 & 0.36 \pm 0.07  & 3.78 \times 3.01 & 12.605 & 1.66 \pm 0.32 &  & \rm NOEMA &  \rm this \; work\\
\rm CO(9-8) &  6.0380 \pm 0.0003 & 202 \pm 32 & 0.27 \pm 0.04  & 2.91 \times 1.57 & 17.768 & 1.40 \pm 0.21 &  & \rm NOEMA & \rm this \; work\\
\rm CO(10-9) &  6.0391 \pm 0.0007 & 233 \pm 34 & 0.24 \pm 0.07  & 2.41 \times 1.06 & 17.812 & 1.38 \pm 0.40 &  & \rm NOEMA & \rm this \; work\\
\hline
\rm \textbf{J0129-0035} &&&&&&&&& \\
\rm [C II] &  5.7788 \pm 0.0001 & 206 \pm 9 & 2.15 \pm 0.08  & 0.22 \times 0.16 &  & 19.2 \pm 0.7 &  & \rm ALMA & V20 \\
\rm CO(2-1) & 5.7783 \pm 0.0004 & 195 \pm 41 & 0.036 \pm 0.005  & 0.69 \times 0.64  &  & 0.039 \pm 0.005 &  & \rm VLA & S19 \\
\rm CO(6-5) & 5.7794 \pm 0.0008 & 283 \pm 87 & 0.37 \pm 0.07  & \sim 5 \times 5 &  & 1.20 \pm 0.23 &  & \rm PdBI & W11 \\
\rm CO(6-5) &  5.7785 \pm 0.0002 & 183 \pm 16 & 0.18 \pm 0.03  & 0.45 \times 0.34 & 71.912 & 0.59 \pm 0.09 &  & \rm ALMA & T24b\\
\rm CO(8-7) &  5.7800 \pm 0.0014 & 283 \pm 187 & 0.21 \pm 0.09  & 4.73 \times 2.46 & -139.37 & 0.91 \pm 0.39 &  & \rm NOEMA &  \rm this \; work\\
\rm CO(9-8) &  5.7780 \pm 0.0005 & 205 \pm 53 & 0.21 \pm 0.05  & 3.43 \times 2.54 & 28.543 & 1.02 \pm 0.24 &  & \rm NOEMA &  \rm this \; work\\
\rm CO(10-9) & \nodata & \nodata & <0.26 & 3.23 \times 2.52 & -6.577 & \nodata &  & \rm NOEMA & \rm this \; work \\
\enddata
\tablecomments{Column 1: Line ID; Columns 2–4: Redshift, linewidth in FWHM and line flux. Note that the line flux is calculated through a single Gaussian fit to the line profile; Column 5: Beam size in FWHM; Column 6: Line luminosity and calibration uncertainties are not included in the error bars; Column 7: Facilities; Column 8: References: \citet{Wang2010}(W10), \citet{Wang2011}(W11), \citet{Shao2019}(S19), \citet{Venemans2020}(V20), \citet{Decarli2022}(D22), \citet{Tripodi2024}(T24a), and \citet{Tripodi2024a}(T24b).}
\end{deluxetable*}

\begin{deluxetable}{CCCC}
\tablecaption{Dust Continuum Observations \label{tab: cont}}
\tablewidth{0pt}
\tablehead{
\colhead{wavelength} & \colhead{telescope} & \colhead{J2054-0005} & \colhead{J0129-0035} 
}
\decimalcolnumbers
\startdata
i (\rm mag) & \rm SDSS & 23.30 \pm 0.22^a & 24.52 \pm 0.25^b \\
z (\rm mag) & \rm SDSS & 20.72 \pm 0.09^a & 22.16 \pm 0.11^b \\
J (\rm mag) & \rm SDSS & 19.18 \pm 0.06^a & 21.78 \pm 0.15^b \\
F_{100 \rm \mu m} (\rm mJy) & Herschel/\rm PACS & < 2.7^c & \nodata \\
F_{160 \rm \mu m} (\rm mJy) & Herschel/\rm PACS & 9.8 \pm 1.3^c & \nodata \\
F_{250 \rm \mu m} (\rm mJy) & Herschel/\rm SPIRE & 15.2 \pm 5.4^c & < 12.2^d \\
F_{350 \rm \mu m} (\rm mJy) & Herschel/\rm SPIRE & 12.0 \pm 4.9^c & < 11.4^d \\
F_{444 \rm \mu m} (\rm mJy) & \rm ALMA & 9.87 \pm 0.94^l & \nodata \\
F_{500 \rm \mu m} (\rm mJy) & Herschel/\rm SPIRE & < 19.5^c & <15.2^d \\
F_{488 \rm GHz} (\rm mJy) & \rm ALMA & 10.35 \pm 0.15^h & \nodata \\
F_{346 \rm GHz} (\rm mJy) & \rm ALMA & 5.723 \pm 0.009^j & \nodata \\
F_{280 \rm GHz} (\rm mJy) & \rm ALMA & \nodata &  2.61 \pm 0.06^i \\
F_{270 \rm GHz} (\rm mJy) & \rm ALMA & 3.15 \pm 0.10^i & \nodata \\
F_{262 \rm GHz} (\rm mJy) & \rm ALMA & 2.98 \pm 0.05^e & \nodata \\
F_{287 \rm GHz} (\rm mJy) & \rm ALMA & \nodata & 2.57 \pm 0.06^e \\
F_{250 \rm GHz} (\rm mJy) & \rm IRAM & 2.38 \pm 0.53^f & 2.37 \pm 0.49^g \\
F_{166.5 \rm GHz} (\rm \mu Jy) & \rm NOEMA & \nodata & 602 \pm 35 \\
F_{161.0 \rm GHz} (\rm \mu Jy) & \rm NOEMA & 542 \pm 24 & \nodata \\
F_{153.0 \rm GHz} (\rm \mu Jy) & \rm NOEMA & \nodata & 494 \pm 37 \\
F_{147.7 \rm GHz} (\rm \mu Jy) & \rm NOEMA & 438 \pm 23 & \nodata \\
F_{134.7 \rm GHz} (\rm \mu Jy) & \rm NOEMA & \nodata & 285 \pm 21  \\
F_{133.3 \rm GHz} (\rm \mu Jy) & \rm NOEMA & 270 \pm 19 &  \nodata \\
F_{102 \rm GHz} (\rm \mu Jy) & \rm IRAM & \nodata &  140 \pm 40^g \\
F_{92.3 \rm GHz} (\rm \mu Jy) & \rm ALMA & 82 \pm 9^k &  \nodata \\
\enddata
\tablecomments{($a$) \citet{Jiang2008}, ($b$) \citet{Jiang2009}, ($c$) \citet{Leipski2014}, ($d$) \citet{Shao2019}, ($e$) \citet{Wang2013}, ($f$) \citet{Wang2008}, ($g$) \citet{Wang2011}, ($h$) \citet{Hashimoto2019}, ($i$) \citet{Venemans2020}, ($j$) \citet{Salak2024}, ($k$) \citet{Tripodi2024}, and ($l$) \citet{Tripodi2024a}.}
\end{deluxetable}

\section{Results} \label{sec: res}
With the new NOEMA observations, we detected the CO (8-7), CO (9-8), and CO (10-9) lines from J2054-0005 in the velocity-integrated maps up to $6\sigma$, $8\sigma$, and $5\sigma$, respectively (Figure \ref{fig: noema_image}). We also detected the CO (8-7) and CO (9-8) lines from J0129-0035 up to $4\sigma$ and $5\sigma$, respectively (Figure \ref{fig: noema_image}). The sources are spatially unresolved in the velocity-integrated intensity maps. The unresolved continuum emission was also detected at high S/N by taking the peak surface brightness value on the continuum map averaging over line-free channels (Figure \ref{fig: noema_image}). We summarize the measurements of the CO lines in Table \ref{tab: line} and the continuum in Table \ref{tab: cont}. 

\subsection{J2054-0005}

This quasar was initially detected in the SDSS stripe 82 with a magnitude of $m_{1450 \, \rm \text{\AA}}=20.60\, \rm mag$ \citep{Jiang2008}. Previous Very Large Array (VLA) observations \citep{Shao2019} detected the CO (2-1) line with a flux of $0.06 \pm 0.01 \, \rm Jy\, km\, s^{-1}$ and an FWHM line width of $270 \pm 47 \, \rm km\, s^{-1}$. The CO (6-5) line detection in previous PdBI observations \citep{Wang2010} reported a line flux of $0.34 \pm 0.07 \, \rm Jy\, km\, s^{-1}$ and an FWHM line width of $356 \pm 106 \, \rm km\, s^{-1}$. However, new ALMA observations \citep{Tripodi2024} show a line flux of $0.29 \pm 0.04 \, \rm Jy\, km\, s^{-1}$ and an FWHM line width of $229 \pm 20 \, \rm km\, s^{-1}$. Considering that the ALMA data has significantly improved sensitivity and more accurate measurement of the line profile, we adopt the ALMA CO (6-5) line measurements for our analysis throughout this paper. \citet{Decarli2022} reported a much broader line width for the CO (7-6) and \ci lines. This inconsistency may be due to the low S/N in their line spectra and the strong noise close to the atmosphere features. We, therefore, exclude the CO (7-6) emission line from the CO SLED analysis in Section \ref{sec: SLEDfit}. The intensity maps of CO(8-7), CO(9-8), and CO(10-9), created by integrating over $\rm 1.5 \times FWHM_{\rm line} $ around the redshifted CO line, are presented in the top-row panels of Figure \ref{fig: noema_image} from left to right. The corresponding continuum images are displayed in the panels of the second row of Figure \ref{fig: noema_image}. We also fitted a single Gaussian profile from the peak position to the line spectra and obtained line fluxes of  $0.36 \pm 0.07 \rm \, Jy\, km\, s^{-1}$, $0.27 \pm 0.04 \, \rm Jy\, km\, s^{-1}$, and $0.24 \pm 0.07 \, \rm Jy\, km\, s^{-1}$ for CO(8-7), CO(9-8), and CO(10-9), respectively. The spectra of CO (8-7) and (10-9) have lower S/N compared to that of the other two transitions and the line widths determined by the fitting suffer large uncertainties. In particular, the line width derived directly from the least square fitting of the CO (8-7) line spectrum yields a value of $331 \pm 62 \, \rm km \, s^{-1}$, which is much larger than that determined with the CO (6-5) and (9-8) emission lines (Table \ref{tab: line}). Therefore, for these two lines, we employ a Bayesian Markov Chain Monte Carlo (MCMC) method and adopt the CO (6-5) line width of $229 \, \rm km \, s^{-1}$ as a prior to fit the line width. To ensure consistency, we also applied MCMC to the CO(9-8) line and found that the result is consistent with the least-square fitting within the uncertainties.

In Figure \ref{fig: noema_image}, the peak positions of the emission are marked with green-yellow crosses, all of which coincide with the optical position \citep[$20^h 54^m 6.4966 \pm 0.0347^s$, $\rm -00^{\circ} 05^{'} 14.500 \pm 0.521^{''} $,][]{Shao2019}. The CO(8-7), CO(9-8), and CO(10-9) emission lines show consistent line widths with those reported for CO(6-5) \citep{Tripodi2024} and  \cii \citep{Venemans2020}.

\subsection{J0129-0035}

This quasar was initially identified in the SDSS stripe 82 with a magnitude of $m_{1450 \, \rm \text{\AA}}=22.28\, \rm mag$ \citep{Jiang2009}. In previous VLA observations \citep{Shao2019}, the CO (2-1) line was detected with a flux of $0.036 \pm 0.005 \, \rm Jy\, km\, s^{-1}$ and an FWHM line width of $195 \pm 41 \, \rm km\, s^{-1}$. The CO (6-5) line detection in previous PdBI observations \citep{Wang2011} reported a line flux of $0.37 \pm 0.07 \, \rm Jy\, km\, s^{-1}$ and an FWHM line width of $283 \pm 87 \, \rm km\, s^{-1}$. However, new ALMA observations \citep{Tripodi2024a} detected the CO (6-5) line with a line flux of $0.18 \pm 0.03 \, \rm Jy\, km\, s^{-1}$ and an FWHM line width of $183 \pm 16 \, \rm km\, s^{-1}$. Similar to J2054-0005, we adopt the ALMA measurements for the CO (6-5) line in our analysis throughout this paper. The intensity maps of CO(8-7), CO(9-8), and CO(10-9), which are also created by integrating over a velocity range of $ 1.5 \times \rm FWHM_{\rm line} $ around the redshifted CO line, are showcased in the third-row panels of Figure \ref{fig: noema_image}. The corresponding continuum images are displayed in the bottom-row panels. For the detections, we extracted the line spectra from the peak position and fitted a single Gaussian profile to the line spectra. The measured line fluxes are $0.21 \pm 0.09 \rm \, Jy\, km\, s^{-1}$ for the CO (8-7) line and $0.21 \pm 0.05 \, \rm Jy\, km\, s^{-1}$ for CO (9-8). Note that the S/N of the CO (8-7) line spectra is insufficient for a robust fitting. Thus, we follow a similar formula as we did for J2054-0005, using the ALMA CO (6-5) FWHM line width of 183 km/s as a prior constraint and employing the MCMC method in the fitting. The $3 \sigma$  upper limit for the CO (10-9) line is calculated as $3\sigma_{\rm channel}(\Delta v_{\rm channel} \Delta v_{\rm line})^{1/2}$(in $\rm Jy \, km \, s^{-1}$), where $\Delta v_{\rm line}$ is adopted as the velocity range $275 \, \rm km\, s^{-1}$ from the ALMA CO (6-5) measurement, $\Delta v_{\rm channel}$ is the binned channel width of $74 \, \rm km\, s^{-1}$, and $\sigma_{\rm channel}$ is the corresponding point-source rms noise value in Jy, respectively \citep{Seaquist1995, Wang2010}. This estimates the upper limit to be $< 0.26 \, \rm Jy \, km \, s^{-1}$. For completeness, we also performed an MCMC fit to the CO(9-8) line and confirmed that the result agrees with the least-square fit, falling within the margin of error.

In Figure \ref{fig: noema_image}, the peak positions of the emission lines are marked with black crosses, all of which match the quasar optical position \citep[$01^h 29^m 58.5150 \pm 0.0347^s$, $\rm -00^{\circ} 35^{'} 39.810 \pm 0.521^{''}$,][]{Shao2019}. The CO(8-7) and CO(9-8) emission lines show consistent line widths with those reported for CO(6-5) \citep{Tripodi2024a} and \cii \citep{Venemans2020}.

\section{Analysis and Discussion} \label{sec: dis}
\subsection{Continuum Emission} \label{sec: sed}
To estimate the total infrared luminosity,$L_{\rm TIR}$, and the underlying SFR of the host galaxies, we present our SED fitting results of the two objects in this section. Here, $L_{\rm TIR}$ refers to the luminosity integrated from 8 to 1000 $\rm \mu m$, while $L_{\rm FIR}$ is integrated over the range from 42.5 to 122.5 $\rm \mu m$ \citep{Helou1985}.

Recent studies suggest that the optically thin assumption may not be valid at FIR wavelength in the compact and dusty starburst systems \citep{Simpson2017, Casey2019}. Therefore, we adopt the general dust opacity modified blackbody model (G$\tau$-MBB) in the continuum SED fitting to determine the dust temperature and dust mass \citep{DaCunha2021}. The dust heating from the cosmic microwave background (CMB) at the quasar redshift with a temperature of $T_{\rm CMB}(z)$ is also taken into account \citep{DaCunha2013}. The SED of the thermal dust emission is modeled as: 
\begin{equation}
S_{\nu_{\rm obs}}^{\rm obs} = \frac{\Omega}{(1+z)^3}[B_{\nu}(T_{\rm dust}(z))-B_{\nu}(T_{\rm CMB}(z))](1-e^{-\tau_{\nu}})
\end{equation}
where $\Omega = (1+z)^4 A D_{\rm L}^{-2}$ is the solid angle covered by the source with $A$, and $D_{\rm L}$ is the surface area and luminosity distance of the source, respectively. $B_{\nu}(\rm T)$ is the Planck function, $T_{\rm dust}$ is the dust temperature, $T_{\rm CMB}(z)$ is the CMB temperature at redshift z, and $\tau_{\nu}$ is the dust optical depth. 

The dust optical depth $\tau_{\nu}$ can be expressed as 
\begin{equation}
\tau_{\nu} = \kappa_{\nu}\Sigma_{\rm d} = \frac{M_{\rm dust}}{A}\kappa_0 (\frac{\nu}{\nu_0})^{\beta}
\end{equation}
where $\beta$ is the emissivity index, $M_{\rm dust}$ is the dust mass, and $\kappa_{\nu} = \kappa_0 (\frac{\nu}{\nu_0})^{\beta}$ is the frequency-dependent dust opacity. In this paper, we adopt $\kappa_0 = 0.77\, \rm cm^2 \, g^{-1}$ at $\nu_0 = 353\, \rm GHz$ (i.e., $\lambda_0 = 850\, \rm \mu m$) that is consistent with \citet{DaCunha2008, DaCunha2015, DaCunha2021}. 

While star formation in the quasar host galaxy dominates the FIR dust emission, the central AGN may also contribute to the dust heating to a large scale ($\sim$kpc). For J2054-0005, the Herschel data measures the continuum emission in the rest-frame MIR-to-FIR bands. To subtract the contribution from the AGN, we utilize a simple SED fitting method to fit the NIR/MIR and FIR multi-wavelength flux density measurements using the Python lmfit package. This method adopted a clump torus model (CAT3D) model from \citet{Honig2010}, which models the NIR and MIR contributions from the AGN dust torus.

For J2054-0005, we approached the SED fitting in two steps. First, we separately fit the SED by combining the G$\tau$-MBB model with each torus template from the CAT3D model. The CAT3D model includes a range of dust torus SED templates generated using various parameters, such as inclination angle and opening angle. In this initial step, we freely fit the G$\tau$-MBB parameters $T_{\rm dust}$ and $\beta$ to achieve the best fit for the SED, using the dust continuum size ($0.9 \times 0.7 \, \rm kpc^2$) measured in \citet{Venemans2020} as a non-fitting parameter. We identified the optimal torus template by choosing the one with the smallest $\chi^2$ value. In the second step, we refined the SED fitting by incorporating the chosen torus template alongside the G$\tau$-MBB model to extract the best-fitting parameters. This method, despite its inherent degeneracy and uncertainty, was aimed primarily at quantifying the AGN's relative contribution without needing to fit the torus's specific parameters precisely. Therefore, our free parameters included the torus component scale factor, $k_{\rm t}$, the torus SED shape determined by the selected template, along with the dust temperature $T_{\rm dust}$, dust mass $M_{\rm dust}$, and emissivity index $\beta$.

\begin{figure}[htbp]
\plotone{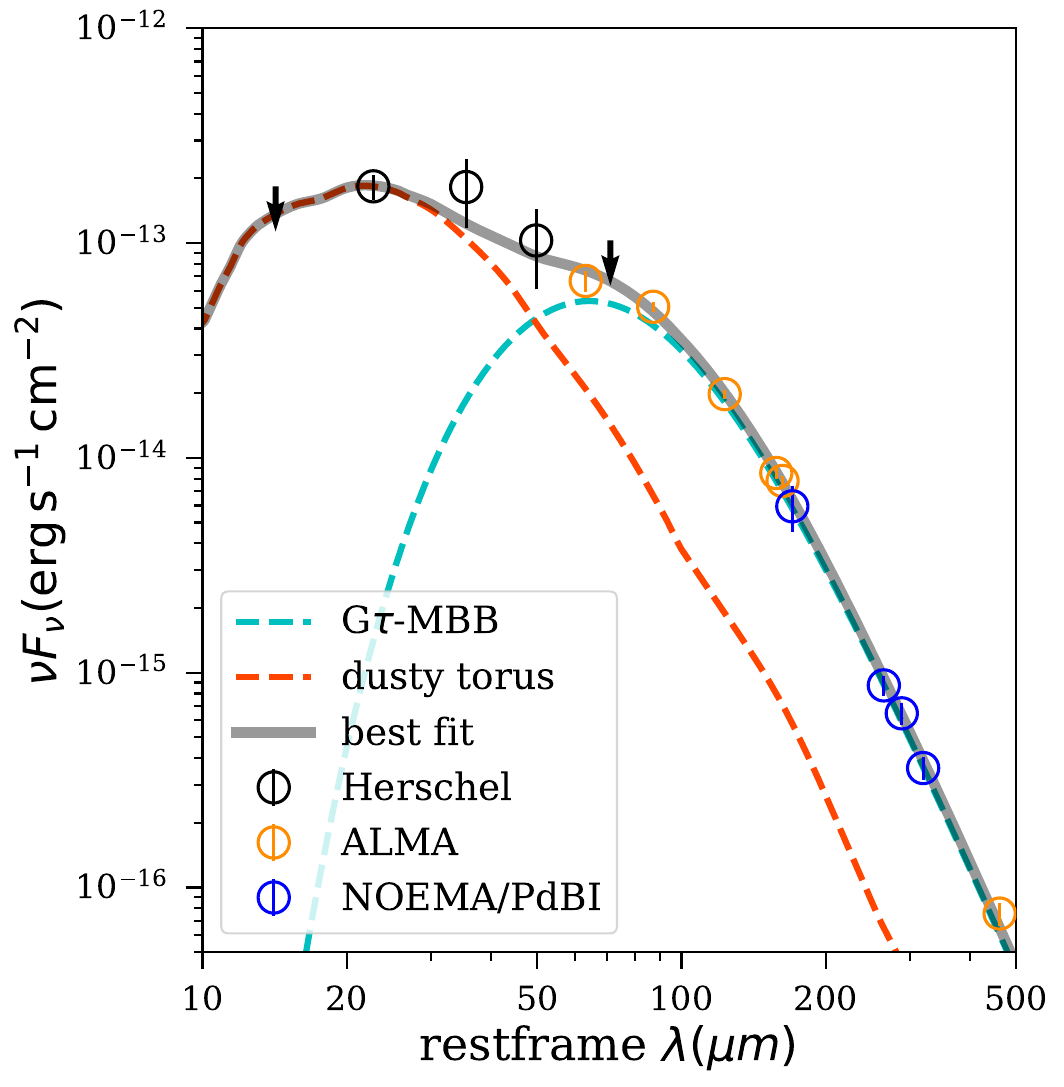}
\caption{The SED fitting results of J2054-0005. Open circles and errors are broadband photometric measurements from Herschel observation \citep[black,][]{Leipski2014}, ALMA observation \citep[orange,][]{Hashimoto2019, Venemans2020, Salak2024, Tripodi2024}, and NOEMA/PdBI observation \citep[blue,][]{Wang2008, Wang2013}. The red-orange dashed line represents the best-fitted CAT3D model and the cyan dashed line shows the best-fitted host galaxy cold dust component.}\label{fig: SED2054}
\end{figure}

Figure \ref{fig: SED2054} displays the best-fit SED for J2054-0005, where the G$\tau$-MBB model yielded parameters of $\beta = 2.17 \pm 0.09$, a dust temperature $T_{\rm dust} = 52 \pm 1 \, \rm K$, and a dust mass $M_{\rm dust} = (1.3 \pm 0.2) \times 10^8 \, M_{\odot}$. \citet{Hashimoto2019} also fitted the SED of J2054-0005, using only four flux density measurements and assuming thin optical depth. They reported a dust temperature $T_{\rm dust} = 50 \pm 2 \, \rm K$ and a dust emissivity index $\beta = 1.8 \pm 0.1$. While their $T_{\rm dust}$ is similar to ours, they found a lower $\beta$, which can be attributed to their assumption of thin optical depth and the limited number of data points used in their fit. Our calculated total infrared luminosity is $L_{\rm TIR} = 6.0 \pm 1.1 \times 10^{12} \, \rm L_{\odot}$, which does not include the torus component heated by AGN. Adopting a Kroupa IMF and the scaling relation ${\rm SFR_{\rm IR}} /({\rm M_{\odot}\, yr^{-1}})= 3.88 \times 10^{-44} L_{\rm TIR}/(\rm erg \, s^{-1}) $ from \citet{Murphy2011}, we estimated the SFR to be $896 \pm 163 \, \rm M_{\odot} \, yr^{-1}$, a value approximately half of that reported by \citet{Hashimoto2019}. This discrepancy primarily arises from our inclusion of the AGN torus in the SED fitting, which \citet{Hashimoto2019} omitted. Additionally, as \citet{Duras2017} suggested, about $50\%$ of the total IR luminosity in AGNs with $L_{\rm bol} > 10^{47} \, \rm erg \, s^{-1}$ is attributed to dust heated by QSOs, a finding consistent with our results.

For J0129-0035, due to the lack of Herschel data at the dust emission peak, $T_{\rm dust}$ has significant uncertainty. We have to assume a SED shape to compute the TIR luminosity. We adopted the graybody with $\beta = 1.6$ and $T_{\rm dust}= 47 \, \rm K$ that was often assumed in the previous optically thin model \citep{Beelen2006, Venemans2020}. We use the size of the dust continuum ($1.2\times 1.0 \, \rm kpc^2$) measured in \citet{Venemans2020} as the galaxy size. Therefore, during our fitting procedure, the free parameter to fit is the dust mass $M_{\rm dust}$. With this assumption, we measured the TIR luminosity to be $L_{\rm TIR} = (4.9 \pm 0.4) \times 10^{12} \rm \, L_{\odot}$, and the dust mass to be  $M_{\rm dust} = (3.9 \pm 0.1)\times 10^8 \rm \, M_{\odot}$. The SFR estimated from $L_{\rm TIR}$ is $733 \pm 59\, \rm  M_{\odot}\, yr^{-1}$. However, due to the unknown AGN contribution, we considered this an upper limit for the real SFR.

\begin{figure}[htbp]
\plotone{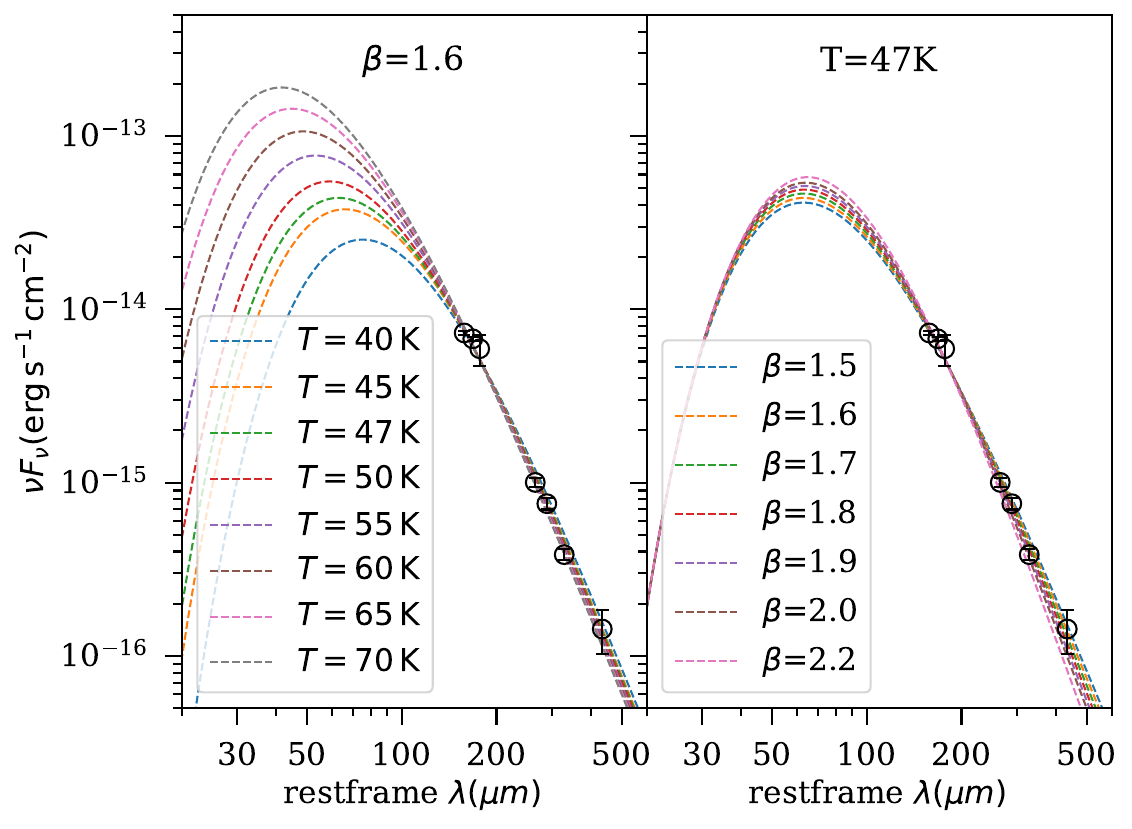}
\caption{The SED fitting results of J0129-0035. The left panel displays the variation of the fitting results with respect to different $T_{\rm dust}$ and fixed $\beta=1.6$. The right panel shows the variation of the fitting results for different $\beta$ and fixed $T_{\rm dust}=47\, \rm K$. }\label{fig: SED0129}
\end{figure}

To address the uncertainty in TIR luminosity and $M_{\rm dust}$ introduced by different assumptions of parameters, we also show the SED models adopting different values of $\beta$ and $T_{\rm dust}$ in Figure \ref{fig: SED0129}. Among these two parameters, the different assumptions of $T_{\rm dust}$ introduce the largest discrepancies in $M_{\rm dust}$ and $L_{\rm TIR}$. In Figure \ref{fig: SED0129}(a), with $T_{\rm dust}$ fixed to $40\rm \, K$, we obtained dust mass of $M_{\rm dust} = (6.1 \pm 0.2)\times 10^8 \rm \, M_{\odot}$, while with $T_{\rm dust} = 70 \rm \, K$, the dust mass is reduced by a factor of 3.7, and $L_{\rm TIR}$ increase by a factor of 7.6. In case we fixed $T_{\rm dust}$ to $40\rm \, K$ and changed $\beta$ from 1.5 to 2.2 (Figure \ref{fig: SED0129}(b)), $M_{\rm dust}$ will decrease by a factor of 3.0 while $L_{\rm TIR}$ moderately increases by a factor of 1.3. 

\subsection{Radiative Transfer Analysis of the CO SLED}\label{sec: SLEDfit}

We further investigate the physical conditions and excitation mechanism of the molecular gas in the quasar host galaxies based on the detections of the CO transitions. We here assume that radiative processes dominate the ISM excitation over non-radiative mechanisms and compare our observed CO SLEDs of J2054-0005 and J0129-0035 with the predictions from PDR and XDR models. Our analysis employs the CLOUDY spectral synthesis code (version c.17.03, \citet{Ferland2017}). Following the formula described in \citet{Pensabene2021}, we have developed an expanded parameter grid comprising $418 \times 2$ models that encompass both PDR and XDR scenarios. These models span a range of total hydrogen density, log($n_{\rm H}\, [\rm cm^{-3}]$), from 2 to 7 (in 19 models with steps of approximately 0.29 dex), and incident radiation field strength (${\rm log}\,  G_0$) from 1 to 7 for PDRs and ${\rm log}\, F_X [\rm erg\, s^{-1}\, cm^{-2}]$ from -2.4 to 2.4 for XDRs. We focus on a specific case with a total hydrogen column density of ${\rm log} (N_{\rm H}/{\rm cm^{-2}}) = 23$ to fully sample the molecular component typically located at $N_{\rm H} > 10^{22} \, {\rm cm^{-2}}$, as observed in giant molecular clouds \citep{McKee1977, Pensabene2021}. We assume that all line emission originates from the neutral or molecular gas phase, disregarding the contribution of HII regions to the \cii emission \citep{Pensabene2021, Decarli2022}. From the CLOUDY simulations, we obtained intensities of the lines of interest and the thermal dust continuum, including CO J=1-20, \cii, \ci, and, $L_{\rm TIR}$.

We then fit the grids described above to the observed CO SLEDs of the two objects by minimizing the $\chi^2$ statistic over the grids of $\G$ and $\nH$:
\begin{equation}
\chi^2 = \sum_{i=1}^{n-m}z_i + \sum_{j=1}^{m}(-2\log\left(\frac{1+\text{erf}(z_j)}{2}\right))
\end{equation}
Here, $n$ represents the measurements of CO lines, including both detections and upper limits, and $m$ denotes the number of upper limits.  The quantity $z_i$ is defined as $z_i = \frac{[y_i - f(x_i; n_{\rm H}, G_0)]^2}{\sigma_i^2}$ \citep{Sawicki2012, Boquien2019}. The quantities $y_i$ and $\sigma_i$ are the flux and the corresponding uncertainty, respectively. The function $f(x_i; n_{\rm H}, G_0)$ represents the model prediction for the line fluxes at each observed transition, and $\text{erf}(z_j)$ is the error function. The modified $\chi^2$ formulation, which includes the error function $\text{erf}(z_j)$, allows us to incorporate upper limit information into the analysis. Specifically, for detected lines, the standard $\chi^2$ term is used. For upper limits, the contribution to $\chi^2$ is weighted by the error function, which effectively penalizes the model less for exceeding upper limits than for deviating from actual detections. This approach helps constrain the model parameters by using all available data, including non-detections.

\subsubsection{Single PDR model}
We first utilize a single PDR component to fit the CO SLEDs for both targets. The results are shown in Figure \ref{fig: SLEDfit}(a) and (d).
 In subsequent fittings of the CO SLED, we normalize the intensities to CO(6-5). Therefore, the fitting parameters include the total hydrogen density of the PDR, denoted as ${\rm log}\, (n_{\rm H,\, PDR})$, and the radiation field strength $\G$ \citep[where $G_0$ is in Habing flux units, i.e. $1.6 \times 10^{-3}\rm \, erg \, s^{-1} \, cm^{-2}$, see][]{Habing1968}. We found that the CO SLEDs of both objects can be fitted with a dense gas component and intense FUV radiation field. The best-fitting parameters are $\nH =6$, $\G =6.5$ for J2054-0005 (Figure \ref{fig: SLEDfit}(a)), and $\nH = 6.0$, $\G = 6.0$) (Figure \ref{fig: SLEDfit}(d)) for J0129-0035.

\subsubsection{double-PDR model} \label{sec: double-pdr}
 As the single-PDR model's extreme physical conditions may oversimplify the actual conditions in PDRs, we now explore a scenario with double PDR components for the CO SLEDs. One PDR component is closer to the typical conditions of the extended molecular ISM and contributes mainly to low-J CO lines. The other PDR component could be denser and give rise to the bright high-J CO lines. As demonstrated by the results of the single PDR, very dense and highly irradiated PDR clumps could likely play a significant role, indistinguishable from XDR regions. We fit the SLEDs of both targets as shown in Figure \ref{fig: SLEDfit}(b) and Figure \ref{fig: SLEDfit}(e) using a combined low-density PDR (PDR I) + dense PDR (PDR II) model. We normalize the CO SLED to CO(6-5), defining the contributions of two components to CO(6-5) luminosity as $k_{\rm PDR\, I}$ and $k_{\rm PDR\, II}$, with $k_{\rm PDR\, I} + k_{\rm PDR\, II} = 1$. The fitting parameters include the total hydrogen density (${\rm log}\, (n_{\rm H,\, PDR \, I})$) and the radiation field strength (${\rm log}\, G_{0, \rm PDR\, I}$) of the PDR I, the total hydrogen density of the PDR II (${\rm log}\, (n_{\rm H,\, PDR\, II})$) and its radiation field strength (${\rm log}\, G_{0, \rm PDR\, II}$), as well as the normalization parameter for the PDR I component ($k_{\rm PDR\, I}$). 
For PDR I, we fix the gas density $\nH =3.4$ and the incident FUV field $\G =1.9$ in the grids as closely described in the local AGNs \citep[$\nH \sim 3.5$, $\G \sim 2$; ][]{Meijerink2007, Meijerink2011, Spinoglio2012}. These specific values are used instead of the exact literature values due to the discrete nature of our grids. They are sufficiently representative of environments from the extended region encompassing the star-forming ring. By adopting this, we assume there is a similar common star-forming ISM component in these quasar host galaxies. The more dense PDR II component's parameters are fitted without constraints. For J2054-0005, the results in Figure \ref{fig: SLEDfit}(b) reproduce the observation of CO SLED with ${\rm log}\, (n_{\rm H,\, PDR \, II}/\rm cm^{-3}) = 6.5$ and ${\rm log}\, G_{0, \rm PDR\, II} = 7.0$. For J0129-0035, the fitting results in Figure \ref{fig: SLEDfit}(e) match the observed CO SLED with ${\rm log}\, (n_{\rm H,\, PDR \, II}/\rm cm^{-3}) = 6.5$ and ${\rm log}\, G_{0, \rm PDR\, II} = 7.0$. The fitting results found that PDR I contributes 56$\%$ and 53$\%$ of the luminosity of the CO (2-1) line for J2054-0005 and J0129-0035, respectively. This indicates that PDR I dominates the CO (2-1) line flux and molecular gas mass, assuming the same conversion factor $\alpha_{\rm CO} \approx 0.8 \, \rm M_{\odot} (K \, km \, s^{-1} \, pc^2)^{-1}$ \citep{Carilli2013}. The fitting results of PDR II for both quasars suggest high gas densities with very strong FUV radiation fields, which are quite close to those from the single-PDR models for both objects. This indicates that the best model points to extreme conditions in PDRs, making the contribution of the PDR I negligible for the mid and high-J CO transitions.

\subsubsection{PDR+XDR model}
We also investigate the possibility of whether an XDR component is involved. We define the contributions of two components to CO(6-5) luminosity as $k_{\rm PDR}$ and $k_{\rm XDR}$, with $k_{\rm PDR} + k_{\rm XDR} = 1$. The fitting parameters include the total hydrogen density of the PDR (${\rm log}\, (n_{\rm H,\, PDR})$) and the radiation field strength ($\G$), the total hydrogen density of the XDR (${\rm log}\, (n_{\rm H,\, XDR})$) and its radiation field strength ($\Fx$), as well as the normalization parameter for the PDR component ($k_{\rm PDR}$). However, the observational data is insufficient to constrain these five parameters of the two components simultaneously. In the double-PDR model (Section \ref{sec: double-pdr}), since both components contribute to the observed FIR dust continuum and fine structure lines, determining their contributions is challenging. As a result, we fixed PDRI based on literature values. For the PDR+XDR model, we assume that the FIR dust continuum and fine structure lines primarily originate from the star-forming  PDR component. This allows us to use the observational line and FIR luminosities to constrain the PDR's physical conditions. Given the limited data and model grid constraints, we fixed the PDR parameters derived from these constraints, instead of fitting a narrower range, to avoid boundary issues.

\begin{figure*}[htbp]
\gridline{\fig{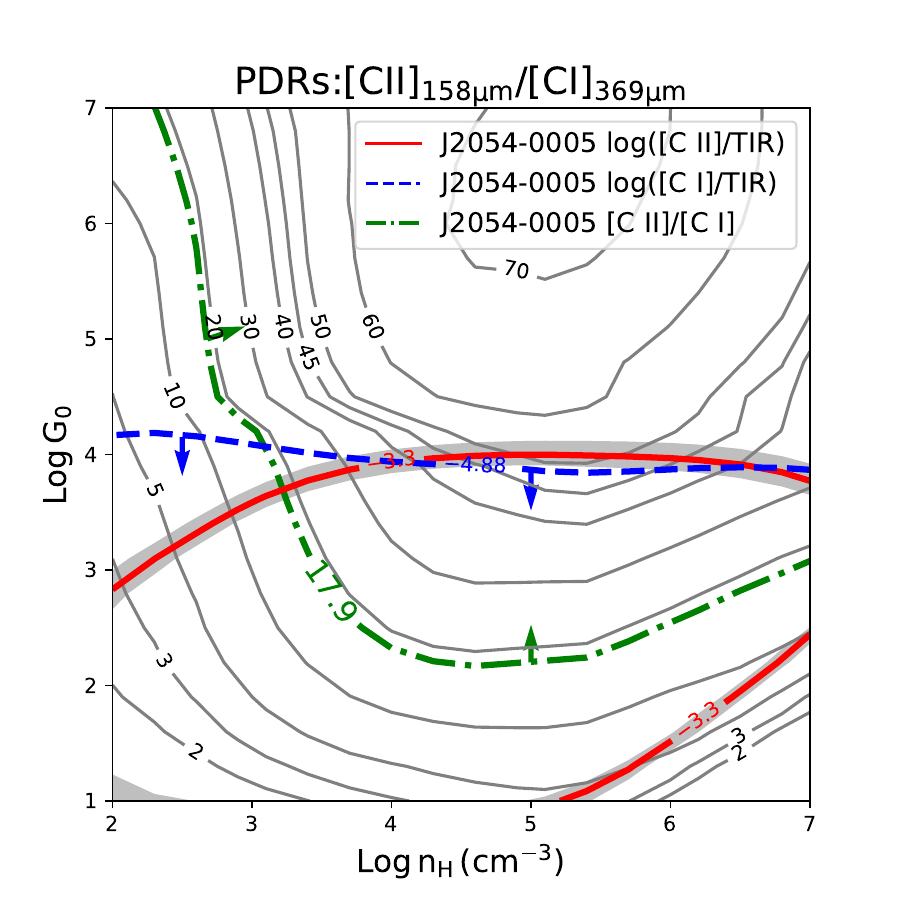}{0.5\textwidth}{(a)}
          \fig{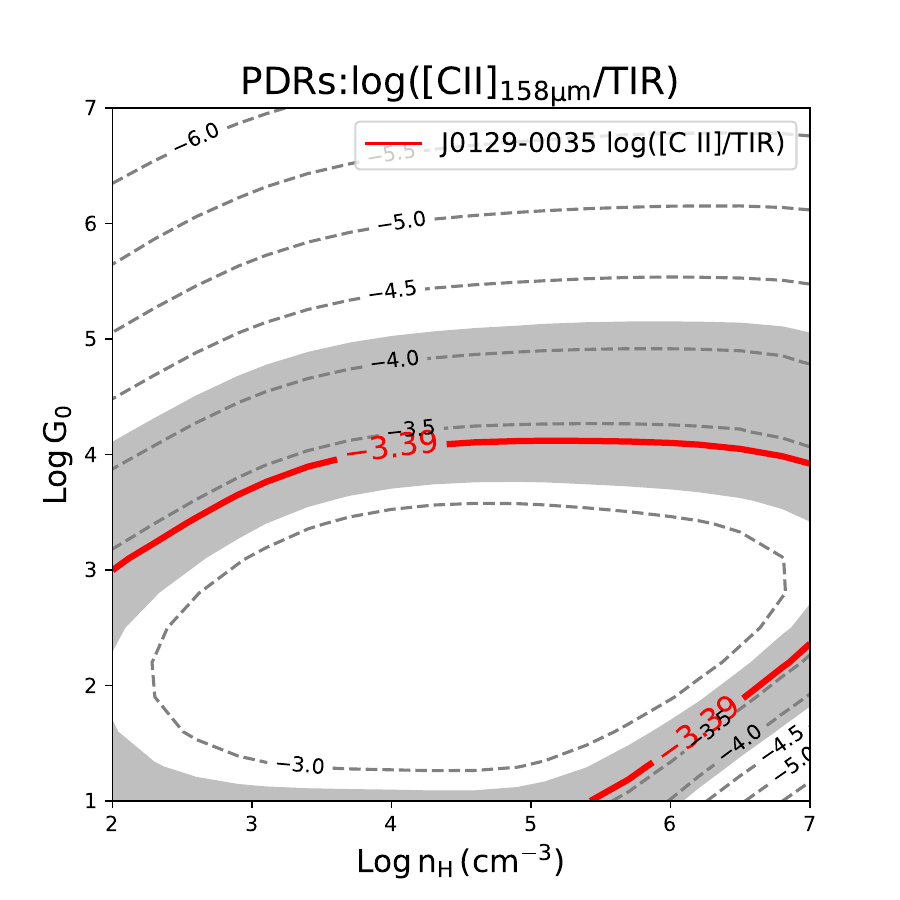}{0.5\textwidth}{(b)}
          }
\caption{(a): Grids of \cii/\ci intensity line ratio as function of $\G$ and $\nH$ in the PDRs. Gray contours indicate model values. The different color lines indicate the line ratio measured in the J2054-0005. The red solid line and gray region represent the \cii/TIR luminosity ratio reported in \citet{Venemans2020}.  The blue dashed line means the upper limit of \ci/TIR luminosity ratio where $3\sigma$ upper limit of \ci in the J2054-0005 was reported in \citet{Decarli2022}. The green dotted line indicates the lower limit of the \cii/\ci luminosity ratio. (b): Grids of \cii/TIR intensity line ratio as function of $\G$ and $\nH$ in the PDRs. Gray contours indicate model values. The red solid line and the line width represent the \cii/TIR luminosity ratio and the uncertainty reported in \citet{Venemans2020}. }\label{fig: diagnosis}
\end{figure*}

In Figure \ref{fig: diagnosis} (a), we utilize the predicted ratio of the \cii, \ci, and total infrared (TIR) luminosity to estimate the physical conditions of PDR in J2054-0005 \citep{Pensabene2021}. Our PDR simulations using CLOUDY also calculate the theoretical ratios between line and continuum intensities. We adopt the measurements of \cii from \citet{Venemans2020} and \tir from Section \ref{sec: sed}, as well as the $3\sigma$ upper limit of \ci from \citet{Decarli2022}. Although these ratios offer poor constraints to the hydrogen density, they effectively constrain the intensity of the radiation field, indicating $\G \lesssim 4.5$ for \cii/\tir and a more stringent limit of $2.5 \lesssim \G \lesssim 4$ for \ci/\tir and \cii/\ci. The lower bound of \cii/\ci becomes more sensitive to the density at lower $n_{\rm H}$ values and increasingly sensitive to $\G$ at $\nH \gtrsim 4$, suggesting the host galaxy resides in a dense and intensely irradiated environment.
Despite large uncertainties and its general insensitivity to the gas density, our diagnostic suggests a moderate radiation field strength, explicitly excluding scenarios with $\G \gtrsim 4.5$. Consequently, our model adopts PDR conditions of $\nH = 3.4$ and $\G = 3.9$. This assumption is based on the intersection of different constraints from Figure \ref{fig: diagnosis} (a), offering a reasonable approximation of the PDR conditions in J2054-0005.

For J0129-0035, we constrain the PDR conditions using the luminosity ratios of \cii and the \tir. The TIR luminosity in this object has very large uncertainties, as discussed in the previous section. \cii has been estimated to arise predominantly from the neutral medium in $z \sim 6$ QSO hosts \citep{Novak2019, Pensabene2021}. We here calculate the upper and lower limits of \cii to TIR ratio adopting the \tir values at $T_{\rm dust}=40 \, \rm K$ and $70 \, \rm K$, which is shown a the gray area in Figure \ref{fig: diagnosis} (b). These constrain the PDR conditions in J0129-0035 to be $\G \lesssim 5.1$. With a similar dust temperature value described in Section \ref{sec: sed}, the \cii/\tir ratio of J0129-0035 is close to that of J2054-0005. Considering this, we also fix the PDR condition of J0129-0035 to $\nH = 3.4$ and $\G = 3.9$ in the PDR+XDR model fitting.

\begin{figure*}[htbp]
\gridline{\fig{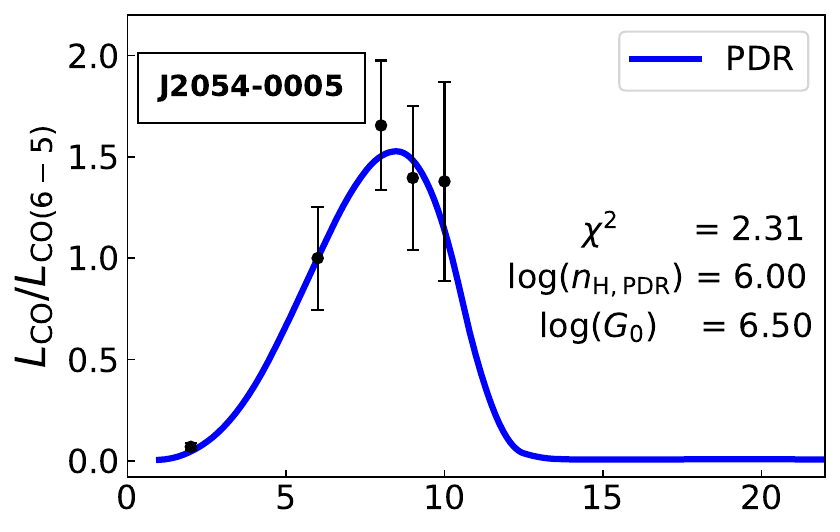}{0.32\textwidth}{(a)}
          \fig{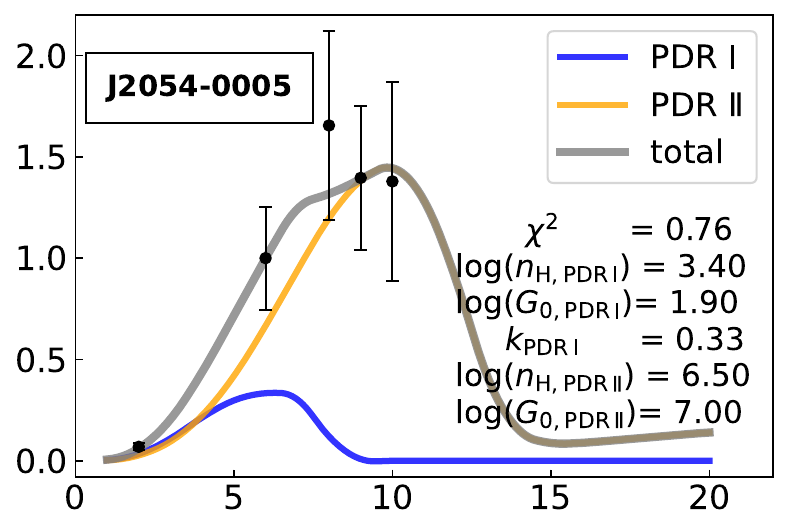}{0.3\textwidth}{(b)} 
          \fig{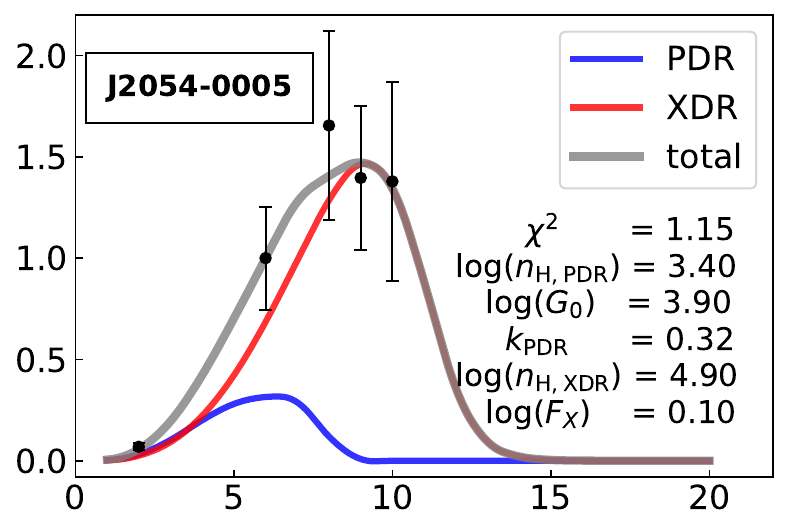}{0.3\textwidth}{(c)}
          }
\vspace{-4mm}        
\gridline{\fig{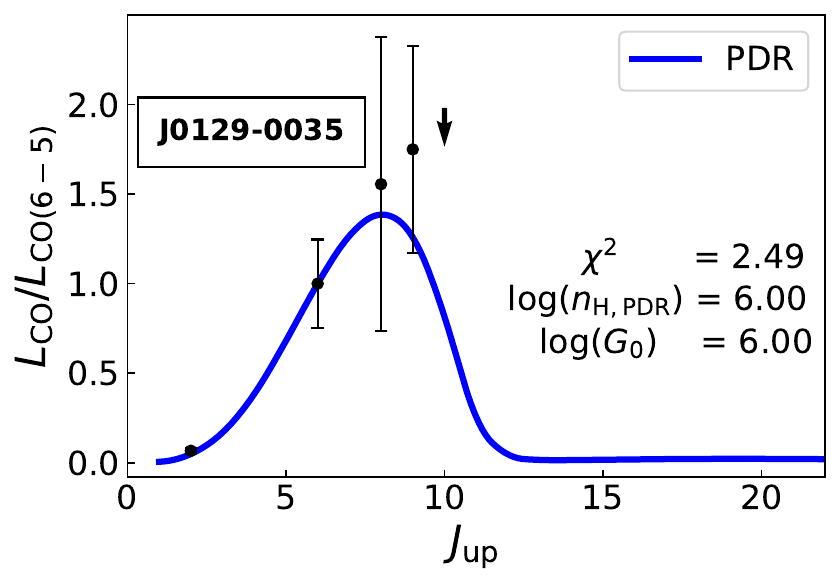}{0.32\textwidth}{(d)}
          \fig{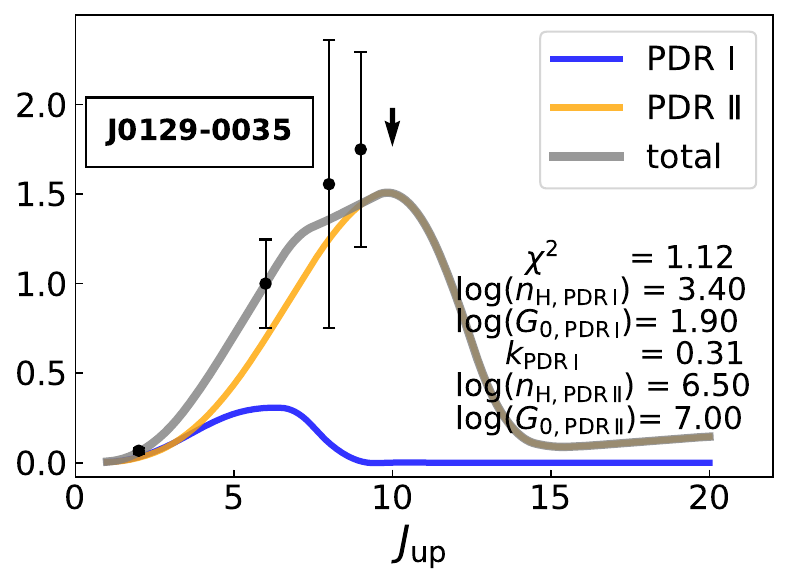}{0.3\textwidth}{(e)}
          \fig{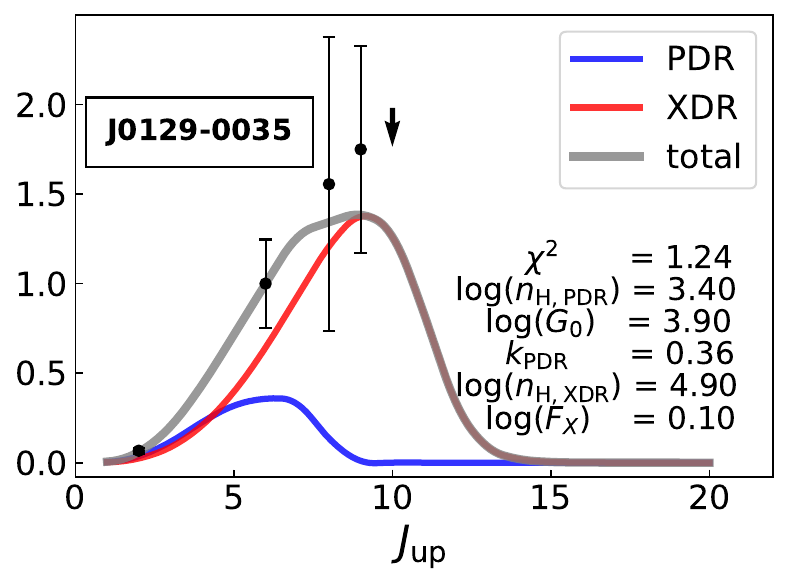}{0.3\textwidth}{(f)}
          }
\caption{CO SLED fitted with PDR and/or XDR models. The black points with error bars represent the CO fluxes and the corresponding uncertainties of J2054-0005 and J0129-0035. The data and/or upper limits for CO(6–5), (8–7), (9–8), and (10–9) are from this work, with CO(6–5) obtained from ALMA observations \citep{Tripodi2024}, and (8–7), (9–8), and (10–9) from NOEMA observations. CO(2–1) data is sourced from \citet{Shao2019}. We determined the optimal parameter combination for the SLED of two sources by fitting the minimum chi-square of the residuals. For J0129-0035, we considered the impact of upper limits on the fitting by incorporating an error function into the residuals. In the upper panel, from left to right, the fits include a single PDR model of J2054-0005, low-density PDR+dense PDR models, and PDR+XDR modeling with the line constraints ($\nH = 3.4$, $\G = 3.9$). The lower panel presents similar cases for J0129-0035, with the line constraints of $\nH = 3.4$, $\G = 3.9$. Blue lines represent the PDR component, red lines indicate the XDR component, and gray lines depict the total contribution from both components.  }\label{fig: SLEDfit}
\end{figure*}

\begin{figure*}[htbp]
\gridline{\fig{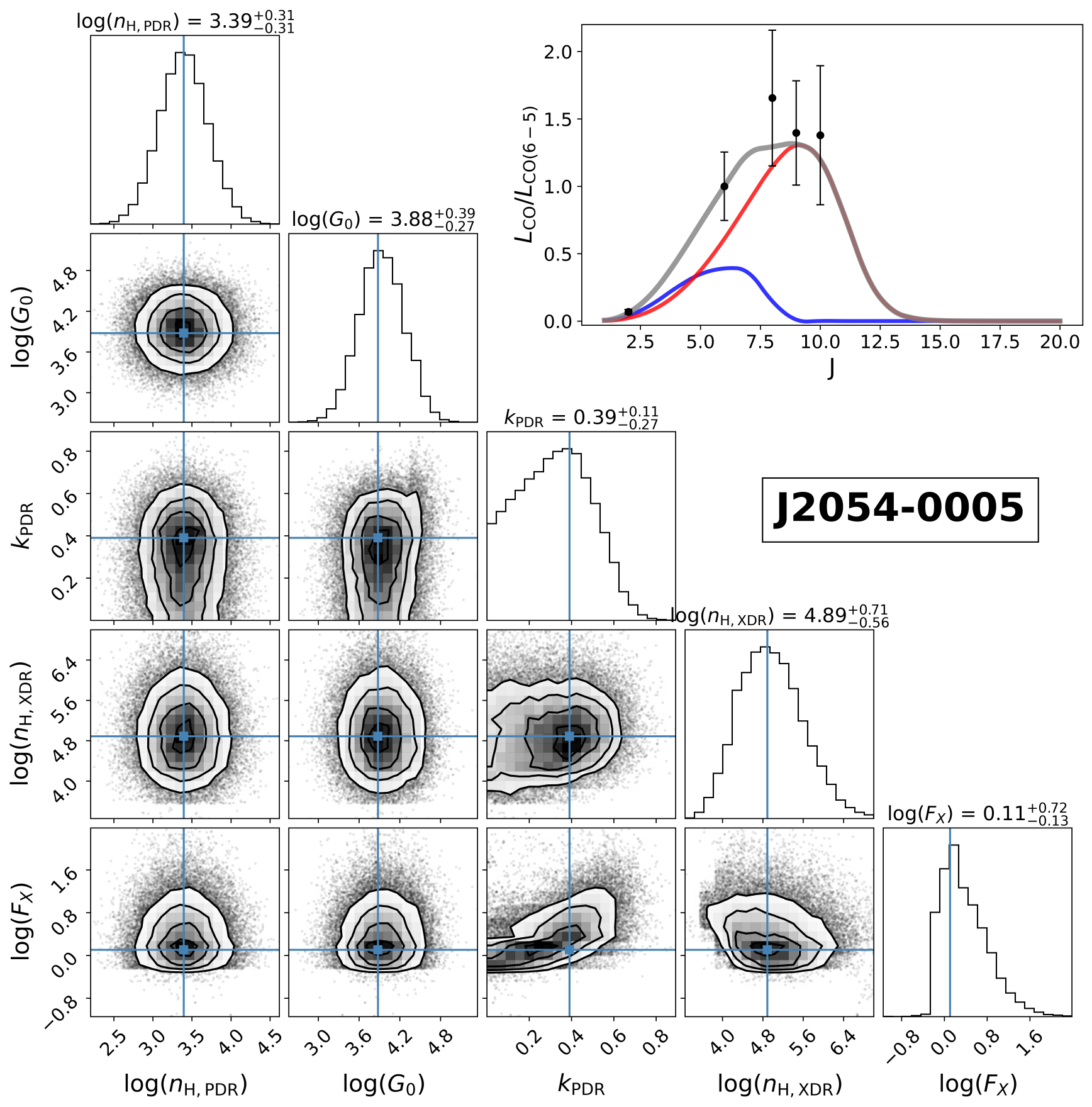}{0.5\textwidth}{(a)}
          \fig{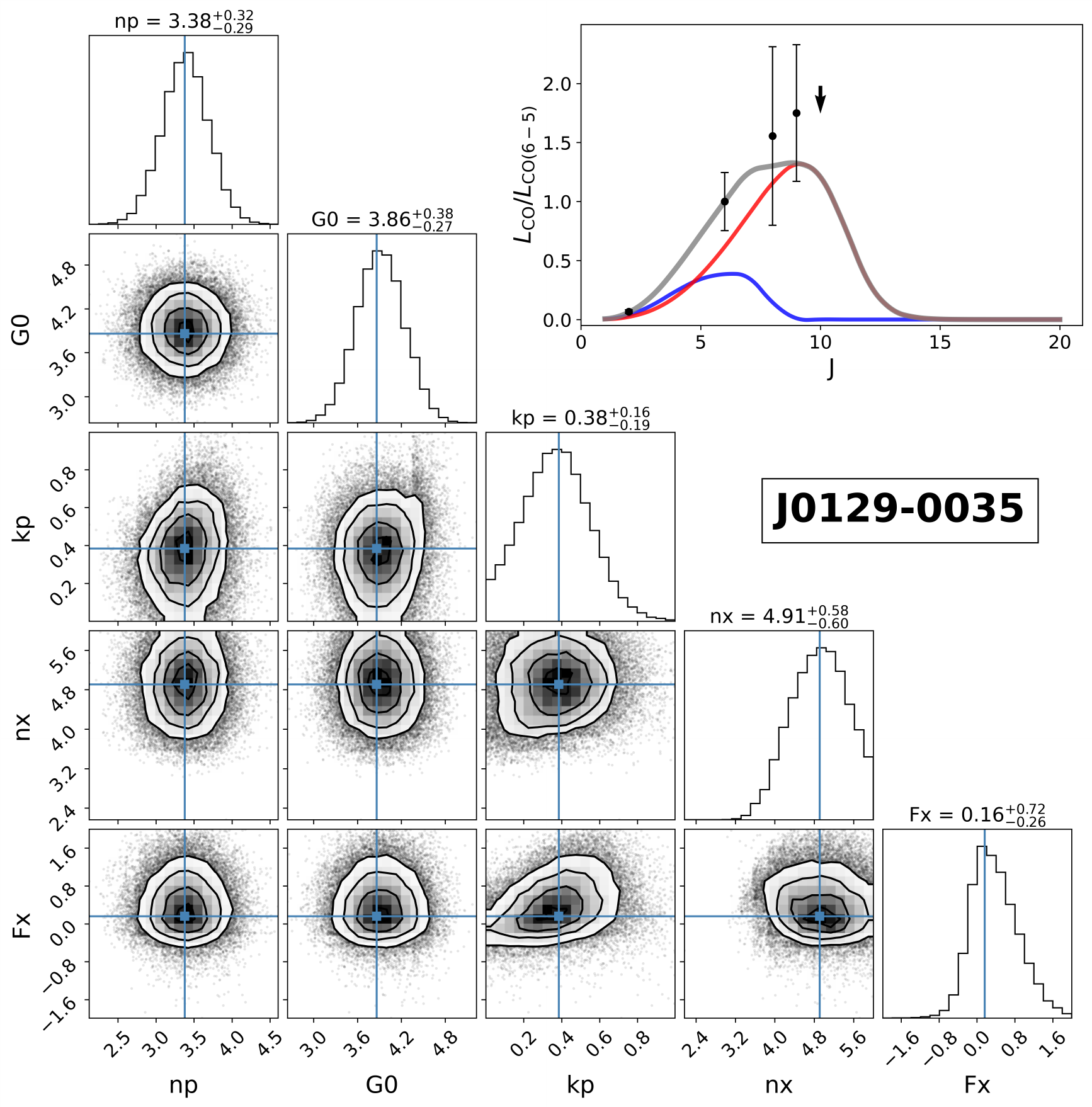}{0.5\textwidth}{(b)}
          }
\caption{ CO SLED fitting results for J2054-0005 (a, left) and J0129-0035 (b, right), employing priors obtained from best-fitting parameters through the least squares method with constrained PDR conditions and predicted line ratios. Blue (PDR) and red (XDR) lines represent the fitting results of the mode of the parameter posterior distribution (cyan). Black dots represent the observed data, and the grey line illustrates the combined PDR+XDR model. }\label{fig: MCMC}
\end{figure*}

With the constraints mentioned above, we fit the CO SLED of both targets as shown in Figure \ref{fig: SLEDfit}(c) and Figure \ref{fig: SLEDfit}(f) using a combined PDR+XDR model. For J2054-0005, the results in Figure \ref{fig: SLEDfit}(c) reproduce the observation of CO SLED with ${\rm log}\, (n_{\rm H,\, PDR}/\rm cm^{-3}) = 3.4$, $\G = 3.9$, ${\rm log}\, (n_{\rm H, \, XDR}/\rm cm^{-3}) = 4.9$, and $\Fx =0.1$. It exhibits a pattern where the excitation at low-J CO transitions is dominated by the PDR component, while the mid to high-J excitations are predominantly driven by the XDR. To estimate the uncertainty of the best-fitting parameters, we employed a Bayesian MCMC fitting algorithm, incorporating Gaussian distribution priors of (${\rm log}\, (n_{\rm H, \, PDR}/\rm cm^{-3}) = 3.1$, $\G = 3.9$, $k_{\rm PDR} = 0.32$, ${\rm log}\, (n_{\rm H, \, XDR}/\rm cm^{-3}) = 4.9$ and $\Fx=0.10$, with $\sigma=0.1$ for ${\rm log}\, (n_{\rm H, \, PDR}/\rm cm^{-3})$ and $\G$, while $\sigma=0.5$ for the others), as predicted by the constrained minimum $\chi^2$ fitting results using line ratio constraints. The best-fitting results are displayed in Figure \ref{fig: MCMC}(a), yielding a derived gas density of $\nH =3.39_{-0.31}^{+0.31}$ and radiation field strength $\G= 3.88_{-0.27}^{+0.34}$, as well as an XDR model with a gas density of $\nH =4.93_{-0.61}^{+0.67}$ and radiation field strength $\Fx = 0.11_{-0.13}^{+0.72}$. 

For J0129-0035, as shown in Figure \ref{fig: SLEDfit}(f), the model also successfully reproduces the observed data with ${\rm log}\, (n_{\rm H,\, PDR}/\rm cm^{-3}) = 3.4$, $\G = 3.9$, ${\rm log}\, (n_{\rm H, \, XDR}/\rm cm^{-3}) = 4.9$ and $\Fx =0.1$. We employed a similar Bayesian MCMC fitting algorithm, taking into account the upper limit of the CO (10-9) line, to fit the combined PDR+XDR model with a Gaussian distribution prior of (${\rm log}\, (n_{\rm H,\, PDR}/\rm cm^{-3}) = 3.4$, $\G = 3.9$, $k_{\rm PDR} = 0.36$, ${\rm log}\, (n_{\rm H, \, XDR}/\rm cm^{-3}) = 4.9$ and $\Fx =0.1$, with $\sigma =0.1$ for ${\rm log}\, (n_{\rm H, \, PDR}/\rm cm^{-3})$ and $\G$, while $\sigma =0.5$ for the others), as predicted by the constrained minimum $\chi^2$ fitting results using line ratio constraints. The optimal fitting results are presented in Figure \ref{fig: MCMC}(b), yielding a derived gas density of $\nH =3.38_{-0.29}^{+0.32}$ and radiation field strength $\G = 3.86_{-0.27}^{+0.38}$, along with an XDR model featuring a gas density of $\nH =4.91_{-0.60}^{+0.58}$ and radiation field strength $\Fx= 0.16_{-0.26}^{+0.72}$. These results align with those obtained from the constrained minimum $\chi^2$ fitting methods.

The best-fitting CO SLED models of the two sources allow us to estimate molecular gas masses and the relative impact of PDR vs. XDR. Assuming a conversion factor $\alpha_{\rm CO} \approx 0.8 \, \rm M_{\odot} (K \, km \, s^{-1} \, pc^2)^{-1}$ and the molecular gas mass $M_{\rm H_2} = \alpha_{\rm CO} L_{\rm CO(1-0)}$ \citep{Carilli2013}, we estimate the total CO(1-0) luminosities to be $5.7 \pm 1.1 \times 10^6 \, \rm L_{\odot}$ and $3.5 \pm 0.5 \times 10^6 \, \rm L_{\odot}$, and the gas masses to be $9.2 \pm 1.8 \times 10^9 \, \rm M_{\odot}$ and $5.7 \pm 0.9 \times 10^9 \, \rm M_{\odot}$ for J2054-0005 and J0129-0035, respectively. We estimate the molecular gas mass of the XDR component to be about $3.9 \pm 0.8 \times 10^9 \, \rm M_{\odot}$ and $2.2 \pm 0.3 \times 10^9 \, \rm M_{\odot}$, contributing approximately 42\% and 39\% of the total molecular gas mass for J2054-0005 and J0129-0035, respectively. 

For both sources, we found that the CO SLEDs can be modeled with a PDR component with very high gas density and intense FUV radiation field ($ n_{\rm H} \sim 10^6 \, \rm cm^{-3}$, $G_0 \gtrsim 10^6$). However, the diagnostics from the fine structure lines prefer PDR components with lower $n_{\rm H}$ and $G_0$ values, which motivates us to consider PDR+XDR models. When we use these line ratios to constrain the PDR conditions and employ a PDR+XDR model to fit the CO SLED, it can be well-modeled by a PDR ($ n_{\rm H} \sim 10^3 \, \rm cm^{-3}$, $G_0 \sim 10^4$) and an XDR ($ n_{\rm H} \sim 10^5 \, \rm cm^{-3}$, $F_{\rm X} \sim 1 \, \rm erg\, s^{-1}\, cm^{-2}$). The PDR gas densities are consistent with the typical conditions of the star-forming component in quasar host galaxies from local to high-z. The XDR components have a higher gas density compared to that of the PDR component, which is reasonable as the X-ray could reach the denser interior of the molecular cloud \citep[][Li et al. 2023]{Meijerink2007, Spinoglio2012, Gallerani2014, Pensabene2021}. However, we cannot fully rule out the scenario that the high-$J$ CO lines are from a dense PDR component. Observations of higher $J$ CO transitions ($J>10$) are required to distinguish between XDR and dense PDR models.

\subsection{Complex CO excitation in the starburst quasar hosts at $z \sim 6$ \label{sec: comp}}

\begin{figure*}[htbp]
\gridline{\fig{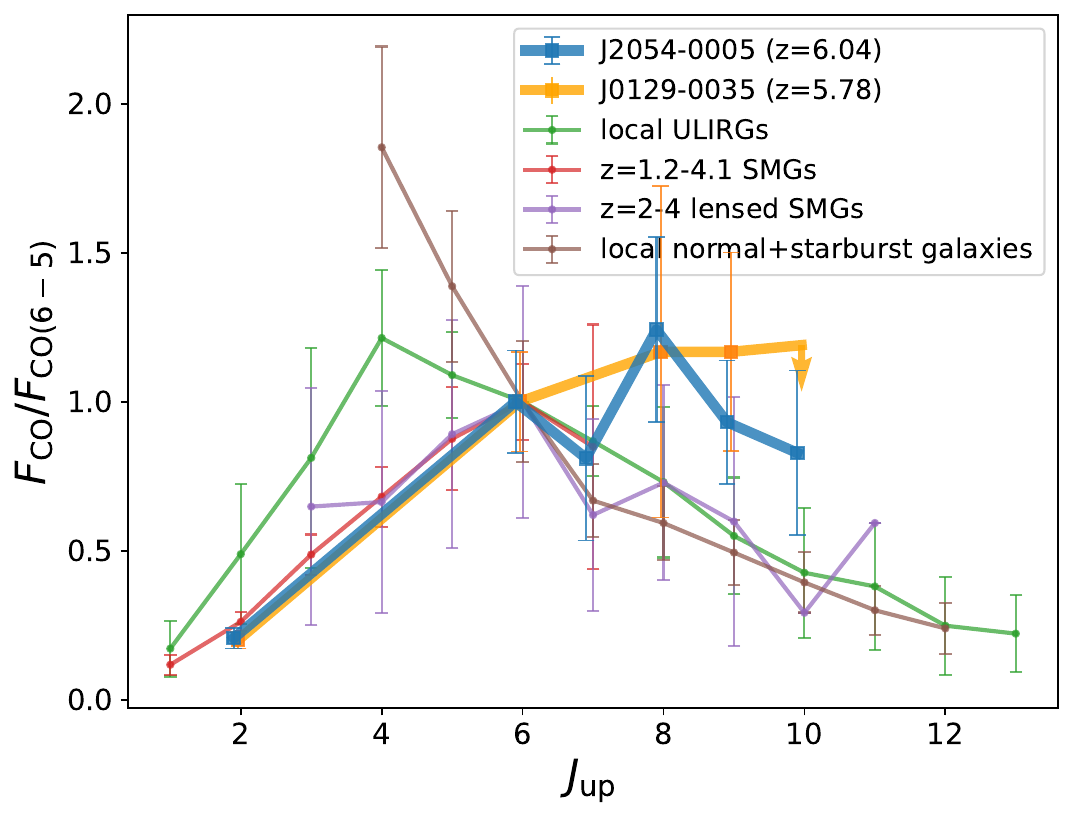}{0.5\textwidth}{(a)}
          \fig{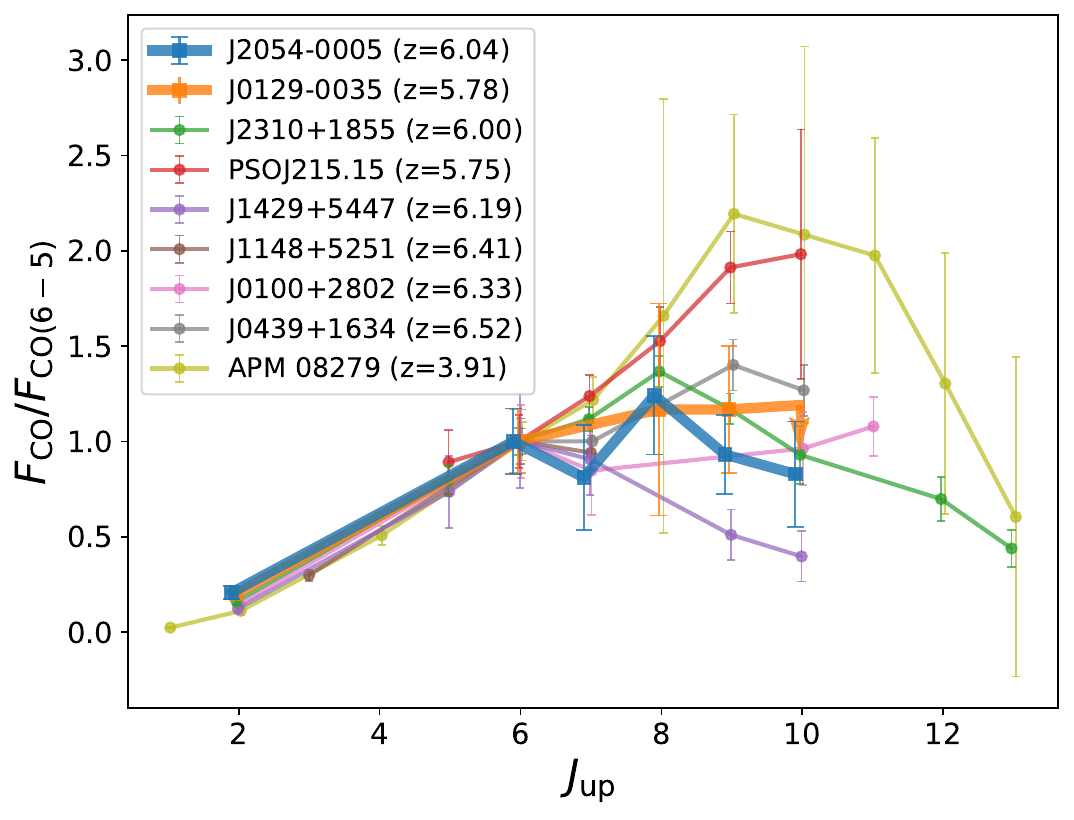}{0.5\textwidth}{(b)}
          }
\caption{ CO SLED normalized to CO(6–5) in normalized unit of $\rm Jy \, km \, s^{-1}$. Left column: J2054-0005 (blue squares) and J0129-0035 (orange squares) in comparison with the mean of four galaxy samples: the local (U)LIRGs \citep[][green]{Rosenberg2015}; local normal + starburst galaxies \citep[][brown]{Liu2015}; $z \sim 1.2-4.1$ SMGs \citep[][red]{Bothwell2013}; and strongly lensed SMGs at $z \sim 2-4$ \citep[][purple]{Yang2017}. Right column: J2054-0005 (blue squares) and J0129-0035 (orange squares) compared with high-z quasars. The plotted quasars are J2310+1855 \citep[][green]{Li2020a}, PSOJ215.15 \citep[][red]{Li2024}, 1429+5447 \citep[][purple]{Li2024}, J1148+5251 \citep[][brown]{Bertoldi2003, Walter2003, Beelen2006, Riechers2009, Gallerani2014}, J0100+2802 \citep[][magenta]{Wang2019}, J0439+1634 \citep[][gray]{Yang2019}, and APM 08279+5255 \citep[][yellow]{Weiss2007, Riechers2009, Bradford2011}. }\label{fig: SLEDsample}
\end{figure*}

The CO SLED is an essential tool for comprehending the physical conditions and molecular gas in galaxies, which is critical for studying the star formation process and AGN power feedback. To investigate the connection between CO excitation and the power source, we compared the CO SLEDs of the two quasars with those of local and high-z starburst galaxies and AGNs. Figure \ref{fig: SLEDsample}(a) displays a comparison with local and high-z galaxy samples, including local ULIRGs samples \citep{Rosenberg2015}, local normal+starburst samples \citep{Liu2015}, (sub)millimeter galaxies at z=1.2-4.1 \citep{Yang2017}, and lensed (sub)millimeter galaxies at z=2-4 \citep{Bothwell2013}. These starburst galaxies show CO excitation, peaking at $J\lesssim 6$ and dropping rapidly at higher J, indicating that the excitation is dominated by PDR. J2054-0005 and J0129-0035 exhibit higher CO excitation with a peak at J=8, which requires PDR with extremely high radiation field strength and density or additional heating mechanisms as described in Section \ref{sec: SLEDfit}. Figure \ref{fig: SLEDsample}(b) presents a comparison to other $z \gtrsim 6$ quasars that have published CO SLEDs with at least four detected transitions (see references in the figure’s caption). These high-z quasars usually exhibit high CO excitation, with CO SLEDs peaked at $J\gtrsim 6$ and bright high-J ($J\gtrsim 9$) CO transitions. Moreover, we did not find a common shape for the CO SLED in these quasar-starburst systems at $z \sim 6$. For example, among the objects shown in Figure \ref{fig: SLEDsample}(b), the intensity ratios between the high-$J$ and mid-$J$ CO transitions vary in a wide range.

We further investigate the correlations between the CO line ratio $r_{\rm 96}$ and the AGN $1450 \, \rm \AA$ absolute magnitude $M_{1450}$, as well as the thermal dust emission powered by host star formation, for a sample of 10 $z \sim 6$ quasars. $r_{ J_2J_1}$ is the brightness temperature ratio between two transitions and is often used to measure the CO excitation, which is expressed as 

\begin{equation}
r_{J_2J_1} = \frac{L^{'}_{{\rm CO} J_2 \rightarrow J_2-1}}{L^{'}_{{\rm CO} J_1 \rightarrow J_1-1}}=
\frac{S_{{\rm CO}\, J_2 \rightarrow J_2-1}}{S_{{\rm CO}\, J_1 \rightarrow J_1-1}}(\frac{J_1}{J_2})^2
\end{equation}

For some sources in the sample, the absence of mid-infrared observations prevents us from constraining the AGN's contribution to $L_{\rm TIR}$ using multi-component SED fitting. However, since the contribution of the AGN torus is primarily concentrated in the 1-40 $\rm \mu m$ range in the rest frame, $L_{\rm FIR}$ is less affected by the AGN contribution compared to $L_{\rm TIR}$. Therefore, in this section, we will adopt the $L_{\rm FIR}$ obtained from a single graybody model fitting to examine the correlation.

\begin{figure*}[htbp]
\gridline{\fig{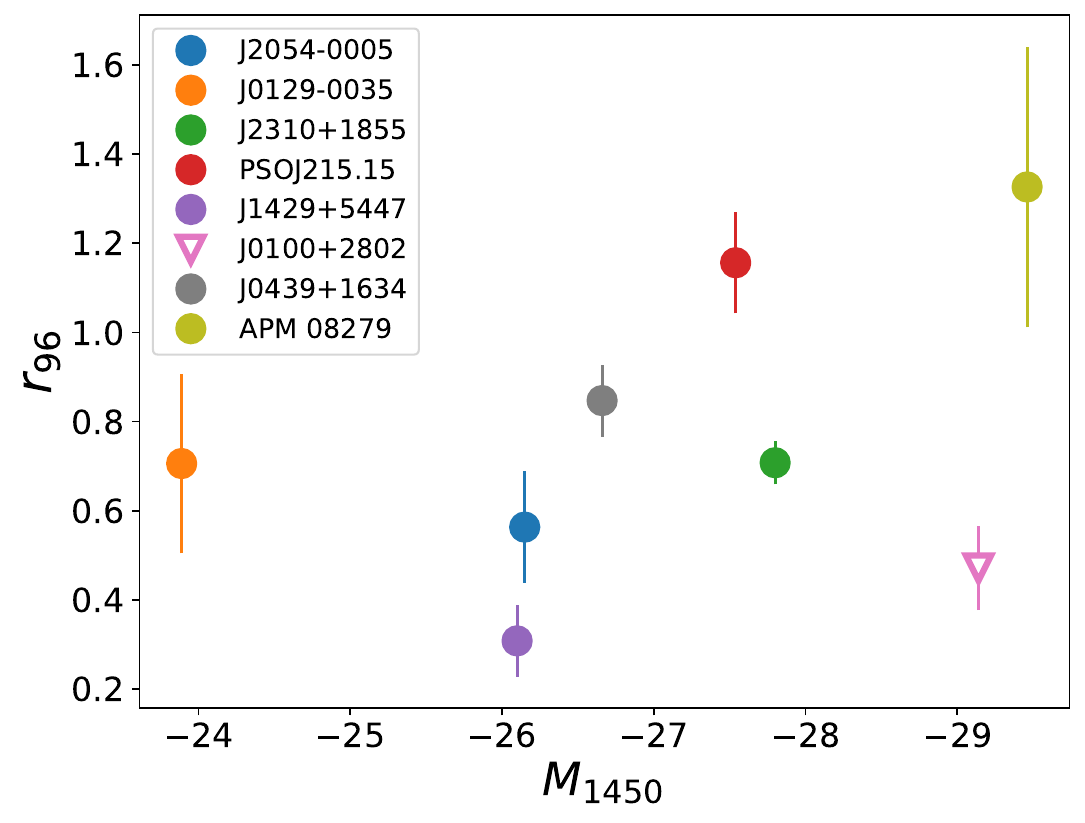}{0.45\textwidth}{(a)}
          \fig{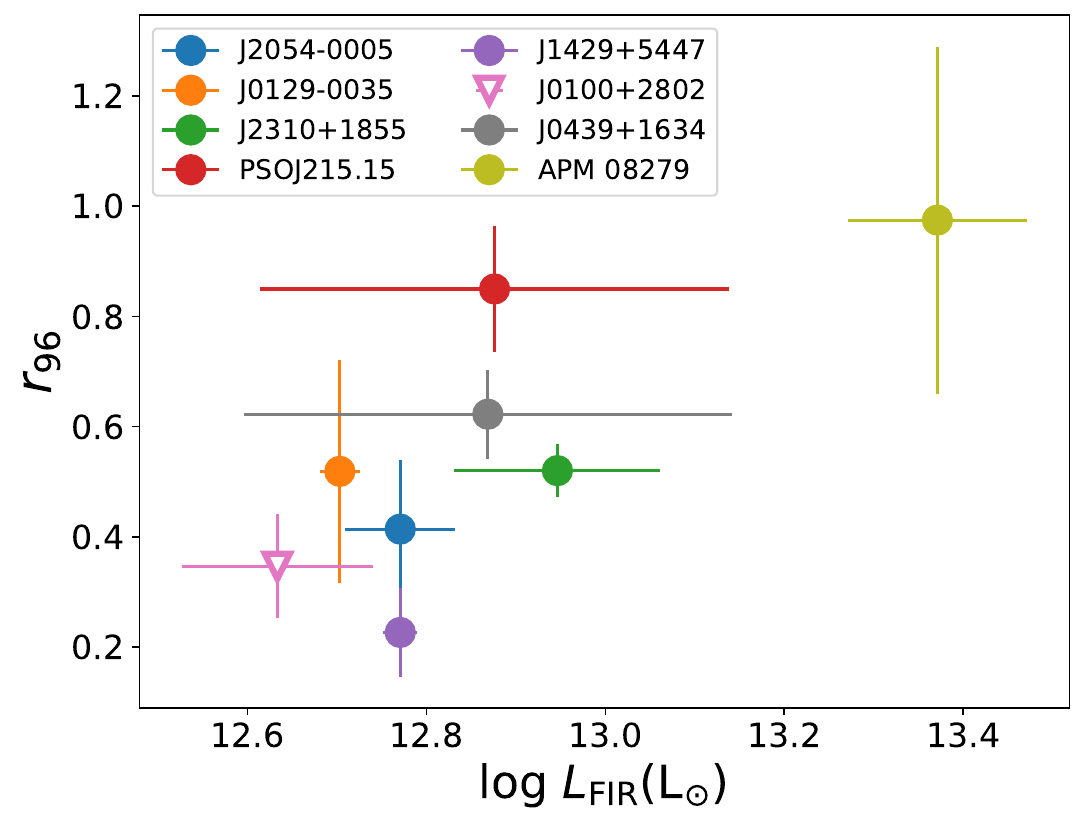}{0.45\textwidth}{(b)}
          }
          
\caption{Relation between the CO brightness temperature ratio $r_{\rm 96}$ and the AGN absolute magnitude $M_{1450}$ (a) and the FIR luminosity $L_{\rm FIR}$ (b). The colored points indicate the value for high-z quasars. The hollow inverted triangle is the bold estimate of $r_{\rm 96}$ derived from $r_{106}$ for J0100+2802.}\label{fig: relation}
\end{figure*}

As shown in Figure \ref{fig: relation}, we employed Weighted Least Squares (WLS) to fit the correlation between $r_{96}$ and both $M_{1450}$ and $L_{\rm FIR}$. There is no significant correlation between $r_{96}$ and $M_{1450}$ with a p-value much greater than 0.05. On the other hand, the relation between $r_{96}$ and $L_{\rm FIR}$ is also very marginal with p=0.071. Moreover, if we exclude the lensed object APM 08279 at z=3.9113, which has the highest luminosity and line ratio, the correlation disappears (p=0.25). This suggests that the correlation is significantly influenced by the extreme case of APM 08279. The lack of correlations between the CO line ratio and both the AGN and star-formation luminosities highlights the complexity of the CO excitation in the starburst AGN hosts. The high-J CO lines from different objects may be dominated by different excitation mechanisms (e.g., PDR vs. XDR), and even in a single object, the CO lines could be dominated by complex gas components with different densities and/or power resources. i.e., The CO (9-8) transition could be contributed by both PDR an XDR, which confounds the connection between $r_{96}$ and the AGN power. Additionally, the X-ray luminosity is more suitable than $M_{1450}$ for investigating XDR. However, only four targets in the comparison sample have detected X-ray luminosity: APM 08279, J0100+2802, J2310+1855, and J1429+5447 \citep{Vito2019, Bertola2022, Migliori2023}. A larger sample of high-z quasars covered by both CO and X-ray measurements is required with further observations to improve the statistics here.

Future observations of higher-J CO ($\rm J \gtrsim 10$) transitions are needed, as the significant role of XDR can be more easily distinguished from PDR in these transitions. Additionally, diagnostics of other FIR lines such as \cii/\ci can help differentiate between the two excitation mechanisms, with the maximum possible ratio for XDR being around 15 \citep[e.g.,][]{Venemans2017, Novak2019, Pensabene2021, Meyer2022}. High-resolution CO observations can map the spatial distribution of these lines, distinguishing between different regions such as star-forming areas and AGN-dominated zones, and thus provide insights into the physical conditions and processes driving the excitation \citep[e.g.,][]{Spinoglio2012}.

\section{Summary} \label{sec: sum}
We have reported NOEMA observations of the CO(8-7), (9-8), and (10-9) lines and the underlying dust continuum in two $z \sim 6$ quasars. These two objects are about 1-2 magnitudes fainter in rest-frame UV ($M_{1450}\sim -24$) compared to the previous $z \sim 6$ sample with CO SLED measurements, which allow us to explore the impact of AGN luminosity on CO excitation in the less luminous quasar population at the epoch of cosmic reionization. We also analyze the physical conditions of the dust and molecular gas through SED fitting and CO SLED modeling, respectively, and compare the CO SLED of the two quasars with local/high-z starburst systems and other $z \sim 6$ quasars. We discuss the relation between CO excitation with AGN and FIR luminosity. The main results are summarized as follows.
\begin{enumerate}
\item We present a detailed analysis of the dust continuum of J2054-0005 at z=6.0379 based on Herschel and ALMA
data and derive the dust temperature $T_{\rm dust} \sim 52 \rm \, K$, the dust mass $M_{\rm dust}\sim 1.3\times 10^8 \, \rm M_{\odot}$, and the total-infrared luminosity $L_{\rm TIR} = 6.0 \times 10^{12} \, \rm L_{\odot}$. We also found that the contribution from AGN is indispensable for the SED fitting in the infrared band. For J0129-0035 at z=5.7787, we fit the SED using G$\tau$-MBB model and derive $M_{\rm dust} \sim 3.9\times 10^8 \, \rm M_{\odot}$ and $L_{\rm TIR} = 4.9 \times 10^{12} \rm L_{\odot}$ with a fixed $T_{\rm dust}=47 \, \rm K$ and $\beta =1.6$.
\item We analyze the CO SLEDs of J2054-0005 and J1029-0035 with available PDR and XDR models. The best-fitting results of both single-PDR and double-PDR models require a gas component with a high gas density ($ n_{\rm H} \sim 10^6 \, \rm cm^{-3}$) and a powerful radiation field ($G_0 \gtrsim 10^6$) to explain the bright high-J CO transitions. However, this is inconsistent with the physical condition derived from the diagnostics of the ratios of the fine structure lines. Adopting the prior of PDR conditions from the line ratios, we found that the best fitting results show a PDR component with $ n_{\rm H} \sim 10^3 \, \rm cm^{-3}$ and $G_0 \sim 10^4$ for the two targets and an XDR component with $ n_{\rm H} \sim 10^5 \, \rm cm^{-3}$ and $F_{\rm X} \sim 1 \, \rm erg\, s^{-1}\, cm^{-2}$. It reveals that the AGN may play a significant role in the CO SLED at $J \gtrsim 6$ even for the optically faint quasar with $M_{1450}\sim -24$. However, we still cannot completely rule out the possibility of an additional extremely dense PDR region.
\item The CO SLEDs of the two targets exhibit higher excitation compared to local/high-z starburst systems and a similar excitation to most other $z \sim 6$ quasars. This result is consistent with the CO SLED fitting results and underlines the AGN's critical role in the CO excitation.
\item By analyzing the relation between $r_{\rm 96}$ with $M_{1450}$ and $F_{\rm FIR}$, we found there is no significant correlation between the CO line ratio $r_{96}$ neither with $M_{1450}$ nor with \tir, highlighting the complexity of the CO excitation. Large samples, other more sensitive gas tracers, and higher angular resolution observations are required to further explore the excitation mechanisms of the molecular gas. 
\end{enumerate}

The study of CO SLEDs in these two sources expands our understanding of quasars at $z \sim 6$ with $M_{1450} \gtrsim -26$. We can further explore the different heating mechanisms with future detections of CO SLEDs, especially high-J CO transitions, in more $z \sim 6$ quasars. Once there are several QSO hosts with consistently modeled CO SLEDs using PDR/XDR, it will be possible to compare gas density and radiation field strength values across different objects to identify any trends with QSO properties derived from other fine-structure lines. 

\begin{acknowledgments}
We acknowledge support from the National Natural Science Foundation of China (NSFC) with grant Nos. 12173002, 11991052, 11721303, and the National Key R $\&$ D Program of China grant No. 2016YFA0400703. This work is based on observations carried out under project number S21DD with the IRAM NOEMA Interferometer. IRAM is supported by INSU/CNRS (France), MPG (Germany) and IGN (Spain). This paper makes use of the following ALMA data: ADS/JAO.ALMA\#2018.1.01289.S. ALMA is a partnership of ESO (representing its member states), NSF (USA) and NINS (Japan), together with NRC (Canada), NSTC and ASIAA (Taiwan), and KASI (Republic of Korea), in cooperation with the Republic of Chile. The Joint ALMA Observatory is operated by ESO, AUI/NRAO and NAOJ. AP acknowledges support from Fondazione Cariplo grant no. 2020-0902. Yali Shao acknowledges support from the National Natural Science Foundation of China (NSFC) with Grant No. 12303001, the Fundamental Research Funds for the Central Universities and the Beihang University grant No. ZG216S2305.
\end{acknowledgments}

\bibliography{main}{}
\bibliographystyle{aasjournal}
\end{document}